\documentclass[useAMS,usenatbib]{mn2e}
\usepackage{subfigure}
\usepackage{graphicx, amssymb}
\usepackage{color}

\title[Star Formation in the First Galaxies]{Star Formation in the First Galaxies I: 
Collapse Delayed by Lyman-Werner Radiation}

\author[C. Safranek-Shrader et al.]
{Chalence~Safranek-Shrader$^{1}$\thanks{E-mail: ctss@astro.as.utexas.edu}, Meghann~Agarwal$^{1}$, Christoph~Federrath$^{2}$, \newauthor Anshu~Dubey$^{3}$, Milo\v s~Milosavljevi\'c$^{1}$, Volker~Bromm$^{1}$\\
$^{1}$Department of Astronomy, University of Texas at Austin, Austin, Texas, USA\\
$^{2}$Monash Centre for Astrophysics (MoCA), School of Mathematical Sciences, Monash University, Vic 3800, Australia\\
$^{3}$FLASH Center for Computational Science, University of Chicago, Chicago, Illinois, USA}

\newcommand{\kelvin}{\mathrm{K}}
\newcommand{\cc}{\mathrm{cm}^{-3}}
\newcommand{\solarmass}{M_{\odot}}
\newcommand{\htwo}{\mathrm{H}_2}

\newcommand{\hminus}{\mathrm{H}^-}
\newcommand{\lya}{\mathrm{Ly}\alpha}
\newcommand{\tcmb}{T_{\mathrm{CMB}}}
\newcommand{\fshhtwo}{f_{\mathrm{shield},\mathrm{H}_2}}
\newcommand{\fshhd}{f_{ \mathrm{shield,HD}}}
\newcommand{\fshhdhtwo}{f_{\mathrm{shield},\mathrm{H}_2,\mathrm{HD}}}
\newcommand{\tff}{t_{\mathrm{ff}}}
\newcommand{\kms}{\mathrm{km}\,\mathrm{s}^{-1}}
\newcommand{\kb}{k_{\mathrm{B}}}
\newcommand{\mh}{m_{\mathrm{H}}}
\newcommand{\ev}{\mathrm{eV}}

\newcommand{\vrms}{v_{\mathrm{rms}}}
\newcommand{\vrad}{v_{\mathrm{rad}}}
\newcommand{\vrot}{v_{\mathrm{rot}}}
\newcommand{\vturb}{v_{\mathrm{turb}}}
\newcommand{\pc}{\mathrm{pc}}
\newcommand{\au}{\mathrm{AU}}
\newcommand{\Mpc}{\mathrm{Mpc}}
\newcommand{\yrs}{\mathrm{yrs}}
\newcommand{\yr}{\mathrm{yr}}
\newcommand{\tvir}{T_{\mathrm{vir}}}
\newcommand{\jlw}{J_{\mathrm{LW},21}}
\newcommand{\cs}{c_{\mathrm{s}}}
\newcommand{\cseff}{c_{\mathrm{s,eff}}}
\newcommand{\few}{\mathrm{few}}

\newcommand{\nhd}{N_{\mathrm{HD}}}
\newcommand{\nhtwo}{N_{\mathrm{H}_2}}
\newcommand{\lj}{L_{\mathrm{J}}}
\newcommand{\fabs}{f_{\mathrm{abs}}}
\newcommand{\fdis}{f_{\mathrm{dis}}}
\newcommand{\tdis}{t_{\mathrm{dis}}}
\newcommand{\mj}{M_{\mathrm{J}}}
\newcommand{\fesc}{f_{\mathrm{esc}}}
\newcommand{\ndis}{\dot{N}_{\mathrm{dis}}}
\newcommand{\sfrff}{\mathrm{SFR}_{\mathrm{ff}}}
\newcommand{\mcold}{M_{\mathrm{cold}}}
\newcommand{\mach}{\mathcal{M}}
\newcommand{\machturb}{\mathcal{M}_{\mathrm{turb}}}
\newcommand{\intensunits}{\mathrm{erg}\,\mathrm{s}^{-1}\,\mathrm{cm}^{-2}\,\mathrm{Hz}^{-1}\,\mathrm{sr}^{-1}}
\newcommand{\omegab}{\Omega_{\mathrm{b}}}
\newcommand{\omegal}{\Omega_{\Lambda}}
\newcommand{\omegam}{\Omega_{\mathrm{m}}}
\newcommand{\ncool}{n_{\mathrm{cool}}}
\newcommand{\tcool}{t_{\mathrm{cool}}}
\newcommand{\tcoolhtwo}{t_{\mathrm{cool,H}_2}}

\newcommand{\kfhminus}{k_{\mathrm{f,H}^-}}
\newcommand{\racc}{r_{\mathrm{acc}}}
\newcommand{\msunperyr}{M_{\odot}\,\mathrm{yr}^{-1}}

\newcommand{\rsoft}{r_{\mathrm{soft}}}

\newcommand{\abunde}{x_{\mathrm{e}}}
\newcommand{\abundhe}{x_{\mathrm{He}}}
\newcommand{\abundd}{x_{\mathrm{D}}}
\newcommand{\abundhd}{x_{\mathrm{HD}}}
\newcommand{\abundhtwo}{x_{\mathrm{H}_2} }
\newcommand{\abundhplus}{x_{\mathrm{H}^+}}
\newcommand{\abundi}{x_i}

\newcommand{\apj}{ApJ}
\newcommand{\mnras}{MNRAS}
\newcommand{\araa}{ARA\&A}
\newcommand{\apjs}{ApJS}
\newcommand{\nat}{Nature}
\newcommand{\apjl}{ApJ}
\newcommand{\aap}{AAP}
\newcommand{\ssr}{Space Sci. Rev.}
\newcommand{\pre}{PRE}
\newcommand{\zap}{Zeitschrift f\"ur Astrophysik}

\begin{document}

\label{firstpage}

\maketitle
\topmargin-1cm

\begin{abstract}

We investigate the process of metal-free star formation in the first galaxies with a high-resolution cosmological simulation. We consider the cosmologically motivated scenario in which a strong molecule-destroying Lyman-Werner (LW) background inhibits effective cooling in low-mass haloes, delaying star formation until the collapse or more massive haloes. Only when molecular hydrogen ($\htwo$) can self-shield from LW radiation, which requires a halo capable of cooling by atomic line emission, will star formation be possible. To follow the formation of multiple gravitationally bound objects, at high gas densities we introduce sink particles which accrete gas directly from the computational grid. We find that in a $1\,\mathrm{Mpc}^3$ (comoving) box, runaway collapse first occurs in a $3\times10^7\,\solarmass$ dark matter halo at $z\approx12$ assuming a background intensity of $J_{21}=100$. Due to a runaway increase in the $\htwo$ abundance and cooling rate, a self-shielding, supersonically turbulent core develops abruptly with $\sim10^4\,\solarmass$ in cold gas available for star formation. We analyze the formation of this self-shielding core, the character of turbulence, and the prospects for star formation. Due to a lack of fragmentation on scales we resolve, we argue that LW-delayed metal-free star formation in atomic cooling haloes is very similar to star formation in primordial minihaloes, although in making this conclusion we ignore internal stellar feedback. Finally, we briefly discuss the detectability of metal-free stellar clusters with the \emph{James Webb Space Telescope}.

\end{abstract}

\begin{keywords}
cosmology: theory --- galaxies: formation --- galaxies:
high-redshift --- stars: formation 
\end{keywords}

\section{Introduction}

Theoretical studies have suggested that the first stars, Population III (Pop III), formed in $\sim10^6\,\solarmass$ dark matter `minihaloes' at redshifts $z\sim15-40$ \citep{CR86,HAI96, TEG97}. Due to a lack of efficient coolants in metal-free gas, Pop III stars are thought to have been more massive than typical stars forming today. The details of their formation process, especially the shape of the Pop III initial mass function (IMF), are still a subject of intense study. Early works suggested these stars were extremely massive, exceeding $100\,\solarmass$, and formed one per minihalo \citep{ABN00, BCL02, Y06}. More recent studies, exploring fragmentation at higher densities \citep{TUR09, STA10, CLA11a, CLA11b, GRE11,GRE12}, or modeling the effects of protostellar feedback \citep{STA11,HO11}, have begun to suggest an IMF extending to lower masses. Detailed simulations that will further constrain the Pop III IMF are needed to assess the role that these stars played in early cosmic milestones, such as chemical enrichment of the IGM, reionzation, and the formation of supermassive black hole seeds \citep[e.g.,][]{BL01,BL04,CF05,BY11}. Furthermore, determining the properties of Pop III stars is necessary for interpreting increasingly detailed observations of high-redshift sources \citep[e.g.,][]{DUN12} and of stellar relics in the local Universe \citep[e.g.,][]{FRE05,BC05,FB10,KAR11}.

Directly observing a chemically pristine stellar population would represent a significant step towards understanding the formation and properties of Pop III stars. However, the chances of detecting individual, high-redshift Pop III stars with the upcoming \emph{James Webb Space Telescope} (JWST) are very low \citep[e.g.,][]{GAR06,GRE09,RYD10}. While a single pair-instability supernova (PISN) from a massive Pop III star may be detectable \citep[e.g.,][]{WA05,HUM11,PAN12,TAN12}, clusters of Pop III stars, if they exist, would present the best opportunity for directly observing a metal-free stellar population \citep{INO11,ZAK11}. It would seem that the haloes capable of hosting these clusters, with virial masses $\sim 10^{7}-10^{8}\,\solarmass$ and often dubbed to be the first galaxies \citep{BYHM09}, would have formed in high-density, biased regions, chemically pre-enriched with Pop III supernova ejecta \citep{TS09,GRE10,WIS12}, precluding the possibility of Pop III star formation. There are scenarios, however, in which these haloes reach the conditions necessary for atomic cooling while still metal-free. For example, if a strong hydrogen molecule dissociating `Lyman-Werner' (LW) radiation background was set up sufficiently early \citep[e.g.,][]{HAI97,MAC01,JOH08} or if Pop III stars ended their lives by collapsing directly to form black holes \citep{HEG03}, the onset of local metal enrichment would have been substantially delayed. Additionally, pockets of metal-free gas could have remained until very low redshifts owing to inhomogeneous metal dispersal \citep{SCA02,FUR05,TOR07,TRE09,ST10,MAI10}. Recently, the detection of metal-free gas clouds at $z\sim 3$ \citep{FU11} has confirmed that regions of space can remain chemically pristine long after reionization.

It has been suggested that two different modes of metal-free star formation occurred in the early Universe. First generation Pop III stars (Pop III.1) formed from initial conditions completely unaffected by previous star formation. Second generation Pop III stars (Pop III.2) formed from gas significantly influenced by the radiative output of previous star formation, but still containing no stellar nucleosynthetic products \citep{OSH08a,MT08,BYHM09}. Cooling by the hydrogen deuteride (HD) molecule is generally thought to differentiate Pop III.1 and Pop III.2 star formation. Unlike $\htwo$, which cannot cool gas below $\sim200\,\kelvin$, HD possesses an intrinsic electric dipole moment and can thus act as an effective cooling agent below $200\,\kelvin$, possibly resulting in stars with lower characteristic masses. 
The abundance of HD can be enhanced in regions with an elevated free electron fraction. These regions can be produced from virialization shocks in haloes with virial temperatures $\tvir> 10^4\,\kelvin$ \citep[e.g.,][]{OH02,GB06} or in the collapse of relic HII regions produced around Pop III.1 stars \citep{FER98,OH03,OSH05,YO07}. For reference, $\tvir$ is related to the virial mass of a halo as \citep[e.g.,][]{BL01}
\begin{equation}
 \tvir \approx  2\times10^4\,\kelvin\left(\frac{\mu}{1.2}\right)\left(\frac{M_{\mathrm{vir}}}{10^8\,\solarmass}\right)^{2/3} \left(\frac{1+z}{10}\right) 
 \label{eq:T_vir}
\end{equation}
where $\mu$ is the mean molecular weight ($\mu=1.2$ for neutral atomic primordial gas), $\mh$ is the mass of a hydrogen atom, $\kb$ is Boltzmann's constant, and $M_{\mathrm{vir}}$ is the total mass contained within the radius in which the average matter density is $18\pi^2\approx178$ times the critical density.
 Theoretical investigations examining the chemistry, cooling, and dynamics of these regions have shown that gas is able to cool to the temperature of the cosmic microwave background ($\tcmb$) as a result of HD cooling. This may result in lower characteristic fragmentation masses \citep[e.g.,][]{JB06}.

One effect that can suppress gaseous collapse, star formation, and metal enrichment in small cosmic haloes is a pervasive UV background. Photons with energies in the range $11.2\,\ev<h\nu<13.6\,\ev$, the LW bands, are capable of photo-dissociating $\htwo$, the key cooling agent in metal-free gas below $10^4\,\kelvin$, and have a very small optical depth in a neutral IGM. Additionally, the photodestruction of $\hminus$, an intermediary in the gas-phase formation of $\htwo$, can also limit the $\htwo$ abundance. Many studies have explored the effect of an UV background on early structure formation \citep{HAI97,CIA00,MAC01,RIC01,MES06,WA07A,YO07,OSH08}. It is accepted that above a certain radiation intensity, $\jlw \approx10^{-1}$, LW radiation delays the collapse and cooling of metal-free gas until the assembly of more massive haloes.\footnote[1]{Here, $\jlw$ denotes the radiation intensity at the centre of the LW bands, $12.4\,\ev$, in units of $10^{-21}\,\intensunits$. This is not to be confused with $J_{21}$, the radiation intensity at the Lyman limit, $13.6\,\ev$, in the same units. In general, $\jlw=\beta\,J_{21}$, where $\beta=3$ for a $10^4\,\kelvin$ blackbody spectrum and $0.9$ for a $10^5\,\kelvin$ spectrum \citep[e.g.,][]{WG111}}. Higher UV background intensities, $\jlw \gtrsim 10$, completely suppress baryonic collapse and cooling in haloes which allow only $\htwo$ cooling. In this regime, significant cooling will not occur until the assembly of larger mass haloes with virial temperatures $\tvir>10^4\,\kelvin$, the atomic cooling threshold. In these haloes, $\lya$ emission will allow the gas to radiate its internal energy and collapse, effectively independent of the radiation background longward of $13.6\,\ev$. 

Studies that have explored the thermodynamical evolution of metal-free gas exposed to a strong UV background \citep{OMU01,OMU08,SS10,WG11,LAT11b} have similarly found that the evolution of metal-free gas undergoing free-fall or isobaric collapse is determined in large part by the background intensity, with the onset of effective $\htwo$ cooling delayed with an increasing $\jlw$. Gas that reaches the atomic cooling threshold is able to collapse isothermally while remaining at $T\sim8000\,\kelvin$ until $\htwo$ forms in sufficient abundance for its cooling rate to exceed the adiabatic heating rate, though this picture depends on the role played by $\lya$ radiation trapping \citep[see][]{LAT11a}. Additionally, there exists a spectrum-dependent critical LW radiation intensity \citep[e.g.,][]{OMU01,SHA10}, $\jlw^{\mathrm{crit}}$, above which $\htwo$ never becomes an effective coolant. In this regime, LW irradiated gas collapses to high density via $\lya$ and $\mathrm{H}^-$ free-bound emission. This evolutionary track has been suggested as a potential mechanism for the formation of supermassive black hole seeds via direct gaseous collapse \citep[e.g.,][]{BL03,BEG06,RH09A,SHA10}. 

The atomic cooling threshold is an appealing criterion for classifying objects as the first galaxies \citep[for a recent review see][]{BY11}. These haloes, with virial masses $\gtrsim5\times10^7\,\solarmass$ at $z\sim10$, distinguish themselves from minihaloes in that in them, metal-free gas can cool and collapse even in the presence of a strong UV radiation field. Additionally, supersonic turbulent gas flows, typically not present in minihaloes, should potentially develop during the assembly of the more massive haloes \citep{WA07,GB08,PRI11}. This turbulence may have influenced the process of star formation in Pop III.2 star forming haloes.

To understand star formation in the first galaxies, it is instructive to take guidance from the better understood case of star formation in the nearby Universe. Overall, the formation of stars is observed to be extremely slow, in the sense that molecular clouds undergoing free-fall collapse should have star formation rates $\sim100$ times higher than the observed rate \citep[e.g.,][]{ZUC74,KT07,EVA09,KV12}. Observational and theoretical work has suggested this inefficiency stems from a number of effects, including supersonic turbulence, protostellar outflows, magnetic fields, and radiative feedback from massive stars. State-of-the-art radiative transfer simulations of star formation including these effects are able to reproduce the IMF and star formation rate in Orion-type Galactic star forming regions \citep[e.g.,][]{KRU12}. While there are undoubtedly many effects which influence star formation, it is becoming accepted that the interplay of supersonic turbulence, gas self-gravity, and protostellar feedback are the key players \citep[e.g.,][]{MK04,ES04,MO07}, setting both the rate of star formation \citep[e.g.,][]{KM05}, and establishing the shape of the stellar IMF \citep[e.g.,][]{PN02,PAD07,HC08,HC11}. With this in mind, we are particularly interested in whether supersonic turbulence plays a role in regulating star formation in the first, metal-free galaxies.

In this work, we present the results of a high-resolution cosmological simulation that follows the assembly of a metal-free, atomically cooling $3\times10^7\,\solarmass$ dark matter halo in an environment exposed to strong LW radiation. We accurately compute the column density of $\htwo$ to properly model the transition where $\htwo$ starts shielding itself from LW radiation, as is necessary for the formation of a cool, dense, baryonic core where star formation can take place. We utilize a spatially adaptive grid to resolve densities up to $n=10^8\,\cc$ and length scales down to $\sim1000\,\au$. Then, we employ sink particles to study the long-term fragmentation tendencies of the gas. Similar studies that focused on LW suppression of $\htwo$ explored smaller values of the radiation intensity, $\jlw\lesssim1$, than we consider here and argued that $\htwo$ self-shielding was not an important effect in the small mass haloes they considered \citep[e.g.,][]{MAC01,OSH08}. Other studies that explored much higher values of $\jlw$, relevant for the theoretical scenario where supermassive black hole seeds form by direct gaseous collapse, only included $\htwo$ self-shielding in an approximate fashion based on purely local estimates of the $\htwo$ column density \citep[e.g.,][]{BL03,SHA10}. This work bridges the gap between these two extremes, exploring the scenario in which collapse in metal-free gas is delayed by LW radiation until the halo reaches the atomic cooling threshold, but with $\jlw$ remaining below $\jlw^{\mathrm{crit}}$ and thus permitting $\htwo$ cooling to play a major thermodynamic role.


We will address two primary questions in this study. First, what is the nature of the fragmentation which occurs in the cold, self-shielding gas? Utilizing sink particles, we can evolve the gas long past the initial gravitational collapse  to times when a significant mass has been accreted by these particles. We can also identify any additional collapsing regions which would represent other potential star forming clumps. In addressing this question, we will comment on the importance of HD cooling which is thought to give rise to enhanced fragmentation, thus producing a distinct population of metal-free stars with lower characteristic stellar masses. Second, is it possible that LW radiation delayed collapse in metal-free atomic cooling haloes can produce clusters of Pop III stars that have luminosities high enough to be detectable and identifiable with the JWST? \citet{ZAK11} showed that clusters of Pop III stars with total stellar masses as low as $\sim10^5\,\solarmass$ should be detectable at $z\approx10$ in deep JWST exposures and that their primordial composition can be ascertained based on simple colour criteria. In a similar study, \citet{INO11} suggested that a star formation rate of a few solar masses per year at $z\sim10$ will be needed for JWST detection. We provide rough estimates for the star formation efficiencies and mass spectra of high-redshift, metal-free stellar populations in atomic cooling haloes and comment on the feasibility of detection.

The outline of the paper is as follows. In Section \ref{sec:numerical_setup} we describe the initial conditions of the simulation and our numerical methodology. In Section \ref{sec:results} we present results of the simulation. In Section \ref{sec:turbulence} we comment on the nature of supersonic turbulence. In Section \ref{sec:fragmentation}, we discuss the trends towards gravitational fragmentation of gas and attempt to predict the properties of the expected starburst. In Section \ref{sec:hd_cooling} we discuss the role of HD cooling. In Section \ref{sec:discussion} we provide further comments on our results, including a discussion on the expected intensity of the LW background, the effect of internal radiative feedback, and the detection prospects of metal-free stellar clusters with the JWST. And finally, we summarize our results and conclude in Section \ref{sec:summary}.

Throughout this paper we assume cosmological parameters consistent with the \emph{Wilkinson Microwave Anisotropy Probe} (WMAP) 7 year results \citep{KOM11}: $\omegal = 0.725$, $\omegab =  0.0458$, $\omegam = 0.275$, $h = 0.704$, $\sigma_8 = 0.810$, and $n_s = 0.967$. Additionally, all quantities will be expressed in physical rather than comoving units unless explicitly stated otherwise.

%
%

\section{Numerical Setup}
\label{sec:numerical_setup}

\subsection{Algorithms and Initial Conditions}
\label{sec:the_simulation}

We use the publicly available adaptive mesh refinement (AMR) code FLASH \citep{FRY00,DUB08,DUB09}, version 3.3, which solves the equations of Eulerian hydrodynamics with the directionally split, piecewise parabolic method of \citet{CW84}. Baryons are represented by a multispecies fluid and dark matter by collisionless, massive particles. The gravitational potential of gas and dark matter is computed with the iterative multigrid Poisson solver of \citet{RIC08}.

We initialize the simulation at $z=146$ in a $1$ $\mathrm{Mpc}^3$ (comoving) box. Cosmological initial conditions were generated with MPGRAFIC \citep{PRU08}, a parallel version of the multiscale Gaussian random field generator GRAFIC \citep{BERT01}. We first run a $128^3$ dark matter only simulation and use the halo finder HOP of \citet{EIS98} to locate the site of the first $10^8\,\solarmass$ dark matter halo in the simulation volume. We then carry out a hierarchical zoom-in procedure to increase the mass resolution around this halo using three separate levels of dark matter refinement to reach a maximum effective resolution of $512^3$ and an effective dark matter particle mass of $230\,\solarmass$ in the target halo itself. We choose the volume of the highest resolution region such that the total mass contained within it is $10^9\,\solarmass$, 10 times the mass of our target halo. We have verified that only the highest resolution dark matter particles are found in our target halo. Given our box size, the expected number of  $10^8\,\solarmass$ dark matter haloes at $z=10$ is of order 10 \citep[e.g.,][]{GB08}.

%
%

\subsection{Resolution and Adaptive Refinement Strategy}
\label{resolution}

In order to capture gaseous collapse to progressively higher densities, we utilize AMR which creates more finely spaced grids (refines) in localized regions. To trigger refinement, we use two separate criteria that are based on local gas properties. We employ a criterion very similar to that used in \citet{WA07} which refines based on baryonic overdensity. In this scheme, the threshold comoving density for refinement is 
\begin{equation}
\rho_{\mathrm{th}} = 3\rho_{\mathrm{b}}2^{3(l-l_i)(1+\phi)} \mbox{ ,}
\label{threshold_density1}
\end{equation}
where $\rho_{\mathrm{b}} = 3H_{0}^2\omegab/8\pi G$ is the comoving baryonic density, $l$ is the current level of refinement, $l_i$ is the initial level of refinement (given our grid resolution, $l_i=5$), and $\phi$ is the Lagrangian refinement factor; $\phi=0$ enforces constant baryonic mass-per-cell while $\phi < 0$ implies that the mass-per-cell decreases with increasing refinement level. We set $\phi = -0.3$ which results in a baryonic mass-per-cell at the highest refinement level ($l_{\mathrm{max}}=22$) of $\approx0.1\,\solarmass$.

A hydrodynamical simulation of a self-gravitating fluid must properly resolve the Jeans length to be physically reliable. For grid-based Eulerian codes, this requirement is expressed by the Truelove criterion \citep{TRU97}, which states that the Jeans length,
\begin{equation}
\lj = \left(\frac{\pi \cs^2}{G\rho}\right)^{1/2}
\label{jeans_length}
\end{equation}
must be resolved by at least 4 grid cell widths to avoid artificial fragmentation. While this criterion was originally formulated for isothermal gas and does not take into account the effect of Hubble expansion, it is commonly utilized in hydrodynamical-cosmological simulations. It would be unnecessary to enforce this criterion across the whole grid, thus we only apply it in the innermost region of the simulation where the most highly refined dark matter particles are present. In this work, we always resolve the Jeans length by at least 12 grid cells and derefine the grid if it is resolved by more than 24. While this is more than sufficient to properly resolve the fragmentation tendencies of gas, it may be insufficient to study the possible small-scale turbulent flow in the gas \citep[see][]{FED11}. 

The FLASH code allows for no explicit force softening beyond local (e.g., one cell wide) cloud-in-cell smearing of the dark matter particle mass density on the computational grid. In our simulation, because of AMR, the grid spacing can become many orders of magnitude smaller than interparticle separation. When this happens, the dark matter particle discreteness can introduce severe artifacts into the calculation of the gravitational potential and cause gas on the computational grid to feel the gravity of individual dark matter particles. To achieve sufficient smoothness in the dark matter particle mass distribution, we have developed an algorithm that spatially smears the dark matter density before it is passed to the Poisson solver. 

For each dark matter particle we compute a smoothing Kernel radius $r_{\mathrm{s}}$ over which the mass of the particle is to be distributed, 
\begin{equation}
r_{\mathrm{s}} = 0.3\,(M_{\mathrm{DM}} /\rho_{\mathrm{b}})^{1/3} \mbox{ ,}
\label{smoothing_radius}
\end{equation}
where $M_{\mathrm{DM}}$ is the dark matter particle mass and $\rho_{\mathrm{b}}$ is the baryonic density of the cell containing the dark matter particle. This choice of smoothing radius ensures that the average dark matter mass per cell contributed by a single dark matter particle inside its smoothing kernel, if multiplied by $\omegab/(\omegam-\omegab)$, is not larger than the baryonic mass inside the particle's host cell. Within the smearing radius, we distribute the particle mass following the quadratic (or `Epanechnikov') kernel $\propto 1-(r/r_{\mathrm{s}})^2$. 

In practice, we achieve the smearing by replacing the dark matter particle with a sufficient number of daughter particles with total mass equal to that of the parent particle and with density approximating the Kernel profile. Given the highly parallel nature of this simulation, this algorithm takes advantage of the FLASH code's capability for moving Lagrangian data between structured, Eulerian blocks \citep[see][]{DUB11,DUB12}. The daughter particle spacing is approximately equal to the cell spacing of the parent particle's computational cell.  Computation of the gravitational potential is performed from the daughter particle density. Computation of the parent particle's acceleration is then carried out in a momentum conserving fashion by summing up the gravitational force over daughter particles and assigning the result to the parent particle.

%
%

\subsection{Chemistry}
\label{chemistry}

Detailed, non-equilibrium chemistry is necessary to properly model the thermodynamic state of cosmological gas flows \citep[for a recent review see][]{G11}. Our chemical model tracks the most relevant chemical species in metal-free gas: $\mathrm{H}$, $\mathrm{H}^-$, $\mathrm{H}^+$, $\mathrm{e}^-$, $\htwo$, $\htwo^+$, $\mathrm{He}$, $\mathrm{He}^+$, $\mathrm{He}^{++}$, $\mathrm{D}$, $\mathrm{D}^+$, and $\mathrm{HD}$. We evolve the species' abundances and the internal energy of the gas by simultaneously integrating $N_{\mathrm{species}}+1$ differential equations with a Bulirsch-Stoer-type, semi-implicit extrapolation mid-point method \citep{BD83}. We set our chemical timestep, which subcycles within the hydrodynamic timestep, as 
$ \Delta t = 0.1\times \mathrm{min}\{ 
n_{e} / |\dot{n}_{e}|,
n_{\htwo} / |\dot{n}_{\htwo}|,
n_{\mathrm{HD}} / |\dot{n}_{\mathrm{HD}}|\}$, 
where $n_i$ is the number density of species $i$.

The initial number density of hydrogen nuclei is given by $n = \bar{\rho}_{b}(z_i) / [\mh  (1.0 + 4.0\,\abundhe)]$, where $\bar{\rho}_{b}(z) = (3H_{0}^2/8\pi G)\, \omegab (1+z)^3$ is the average, physical baryonic matter density at redshift $z$, and $\abundhe= 0.08$ is the primordial number fraction of helium. We set $\abundhtwo= 2\times10^{-6}$, $\abundhplus = 3.8\times10^{-4}$, and $\abundd = 4.3\times10^{-5}$, where $\abundi$ is the number density of species $i$ relative to $n$, the number density of hydrogen nuclei (henceforth the abundance of that species). The abundance of electrons is calculated by enforcing charge neutrality. The initial gas temperature is set by assuming adiabatic cooling due to Hubble expansion after $z\approx200$, when gas and the CMB thermally decoupled.

Properly modeling the formation of $\htwo$ and HD is essential since these molecules are the only low temperature ($<10^4\,\kelvin$) coolants in metal-free gas. In the absence of dust, $\htwo$ forms primarily through the gas-phase reaction mediated by $\mathrm{H}^-$,
\begin{eqnarray}
\mathrm{H} + e^- &\rightarrow&  \mathrm{H}^- + \gamma  \mbox{,}  \nonumber \\
\mathrm{H}^- + \mathrm{H} & \rightarrow&   \htwo + e^-   \mbox{,}
 \label{h2_form}
\end{eqnarray}
which is feasible in primordial gas owing to the residual ionization fraction present after recombination. At higher densities, $\gtrsim10^8\,\cc$, $\htwo$ can also form directly through 3-body reactions \citep{PAL83}, though
since we do not simulate densities this high, 3-body $\htwo$ formation will not be significant here.

The HD molecule can be a significant coolant at temperatures $< 200\,\kelvin$ that can potentially cool the gas to $\tcmb$. In the absence of a LW background, the HD abundance is primarily determined by its main formation pathway
\begin{equation}
\htwo + \mathrm{D}^+ \rightleftharpoons \mathrm{HD} + \mathrm{H}^+ \mbox{,}
\label{eq:hd_form}
\end{equation}
where the $\mathrm{D}^+$ to $\mathrm{H}^+$ abundance ratio is set by the charge exchange reaction
\begin{equation}
\mathrm{H} + \mathrm{D}^+ \rightleftharpoons \mathrm{D} + \mathrm{H}^+\mbox{.}
\label{eq:hplus_dplus_charge_ex}
\end{equation}
Given equilibrium in reactions (\ref{eq:hd_form}) and (\ref{eq:hplus_dplus_charge_ex}), the HD abundance is \citep[e.g.,][]{OMU05}
\begin{equation}
\abundhd \approx 2\, \mathrm{exp}(421 \,\kelvin \,/\, T)\, \abundhtwo\,\abundd.
\label{eq:hd_eq}
\end{equation}
Therefore with HD in chemical (not photodissociation) equilibrium, the HD to $\htwo$ abundance ratio can exceed the cosmological D to H ratio by a large factor when $T\ll421\,\kelvin$. This HD to $\htwo$ fractionation, which is due to the slightly higher binding energy of the HD molecule compared to that of $\htwo$ \citep{SW73}, is a reason why HD is thought to be a significant coolant in Pop III.2 star formation where additional cooling from $\htwo$ can induce an elevated HD abundance.
As we shall show, even in the presence of a strong LW background, Equation (\ref{eq:hd_eq}) describes the HD abundance fairly accurately, although since only a very small fraction of the gas cools down to $T\ll400\,\kelvin$, significant HD to $\htwo$ fractionation does not occur.

%
%

\subsection{Gas Cooling}
\label{gas_cooling}

Atomic and molecular radiative cooling processes have been well studied in astrophysical settings \citep[e.g.,][]{SK87,CEN92,OF06}. For the densities, temperatures, and chemical compositions relevant to this work, the most important cooling mechanisms are $\lya$, or more generally atomic, emission from neutral hydrogen, ro-vibrational emission from molecular hydrogen, and emission from hydrogen deuteride. For completeness, we also include Compton heating and cooling due to electron scattering of CMB photons, H and He recombination and collisional ionization cooling, and free-free emission. We do not consider cooling by metals or dust as we specifically focus on primordial gas.

$\lya$ cooling becomes significant above $\sim 10^{4}\,\kelvin$ when the electron fraction and gas temperature become high enough to excite this transition. This extremely efficient cooling channel can cool the gas to $\sim 8000\,\kelvin$. Ro-vibrational emission from molecular hydrogen can potentially cool the gas further. Being a symmetric molecule, however, $\htwo$ lacks a permanent electric dipole moment and consequently cannot cool the gas below $\sim 200\,\kelvin$. HD, if present, can cool gas to even lower temperatures, $\lesssim100\,\kelvin$, as it does possess an intrinsic dipole moment and more closely spaced energy levels. Nevertheless, an elevated electron fraction is still required for it to form in significant quantities \citep[e.g.,][]{JB06}. We adopt the $\htwo$ cooling rate from \citet{GP98} and the HD cooling rate from \citet{FL02}.

Finally, as the CMB imposes a lower limit on the temperature to which gas can radiatively cool, we adopt an effective cooling rate of the form $\Lambda_{\mathrm{eff}}(T) = \Lambda(T) - \Lambda(\tcmb)$, where $\tcmb = 2.7\,\kelvin\,(1+z)$ and $\Lambda(T)$ is the total volumetric cooling rate not taking into account the CMB. This formulation ensures the gas temperature will not fall below $\tcmb$ unless it does so via adiabatic expansion.

%
%

\subsection{Sink Particles}
\label{sec:sink_particles}

We utilize sink particles to follow the global evolution of the gas after the first baryon dominated region inside a halo undergoes runaway gravitational collapse. Without sink particles, the first collapse would drive the computational mesh to arbitrarily high refinement, which would impose a prohibitively short Courant-Friedrichs-Lewy (CFL) timestep on the entire simulation. Sink particles allow us to self-consistently simulate the formation of multiple gravitationally bound collapsed structures over many free-fall times by accreting mass from the grid and putting an upper limit on the gas density. This computational method was originally introduced in \citet{BBP95} and has been used extensively in numerical simulations, both in AMR \citep[e.g.,][]{KRU04,WA10,FED10,PN11,GIR11} and smoothed particle hydrodynamics \citep[SPH; e.g.,][]{BCL02,BBB03,STA10,GRE11,CLA11a}. 

Our sink particle implementation is identical to the method introduced in \citet{FED10}. Special care is taken to avoid the spurious creation of sink particles that may not represent gravitationally collapsing regions. To this end, we utilize additional checks for the creation of a sink particle in addition to requiring that the gas density in a cell be greater than some threshold $\rho_{\mathrm{thresh}}$. This includes enforcing that the local velocity divergence is negative and that the region of collapse is both gravitationally bound and Jeans unstable. 

Once formed, sink particles are capable of accreting mass directly from the computational grid. Cells with centres within a constant distance, the accretion radius $\racc$, from a sink particle and with a density $\rho>\rho_{\mathrm{thresh}}$ are examined. For these cells, the mass increment $\Delta M=(\rho-\rho_{\mathrm{thresh}})\Delta V$ is calculated, where $\Delta V$ is the cell volume. Provided that the cell velocity is radially directed towards the sink particle and that the mass increment $\Delta M$ is gravitationally bound to the sink particle and the gas within $\racc$, the increment is deducted from the grid and added to the sink particle.

The gravitational interaction between sink particles and gas is computed by direct summation. To avoid extremely large accelerations, we employ cubic spline softening \citep[e.g.,][]{PM07} which decreases the gravitational attraction of gas-sink and sink-sink interactions within $r<\rsoft$, where $r$ is the separation between a cell centre and a sink particle or between two sink particles. 

In the simulation here, we set the sink particle accretion and softening radii to be equal, $\racc=\rsoft= 0.01\,\pc \approx 2000 \,\au$, which is $2.5$ times the grid spacing at the highest level of refinement. Additionally, we set the sink particle creation density threshold to $\rho_{\mathrm{thresh}}=2.2\times10^{-16}\,\mathrm{g}\,\cc$ corresponding to $n=10^8\,\cc$.

%
%

\subsection{Transport of $\htwo$-Dissociating Radiation}
\label{sec:lw_radiation}

The occurrence of a metal-free atomic cooling halo requires previous suppression of Pop III.1 star formation in minihaloes that could have polluted the atomic cooling halo with metals. As previously discussed, this suppression could have been produced by a molecule-dissociating radiative background produced by neighboring star-forming galaxies. From the initial redshift, we impose a constant LW background incident flux with intensity $J_{21} = 100$ (corresponding to $\jlw=90$) onto each of the six faces of the computational box. This ignores periodicity of the domain, which is acceptable given that our target halo is located near the centre of the box. Due to absorption by neutral hydrogen in the IGM, we set $J_{\nu}=0$ shortward of the Lyman limit. We assume the spectral shape of this source is a $10^{5}\,\kelvin$ blackbody, representative of massive, metal-free stars \citep{BK01,SCH02}. While it has been suggested that the photodissociation of $\mathrm{H}^-$ can be crucial in regulating the $\htwo$ abundance, we neglect it as it has been shown to have minimal importance for $10^5\,\kelvin$ blackbody sources \citep[e.g.,][]{OMU08}. However, if Pop III stars were not extremely massive ($\sim100\,\solarmass$) but had more moderate masses ($\sim10\,\solarmass$), the appropriate radiation source temperature would be somewhat lower. We note, though, that the primary consequence of including $\mathrm{H}^-$ photodissociation in this context would be a shift in the density at which $\htwo$ cooling becomes effective, and would not significantly alter our overall results.

\begin{figure}
\includegraphics[scale=0.5, clip, trim = 20 5 5 5  ]{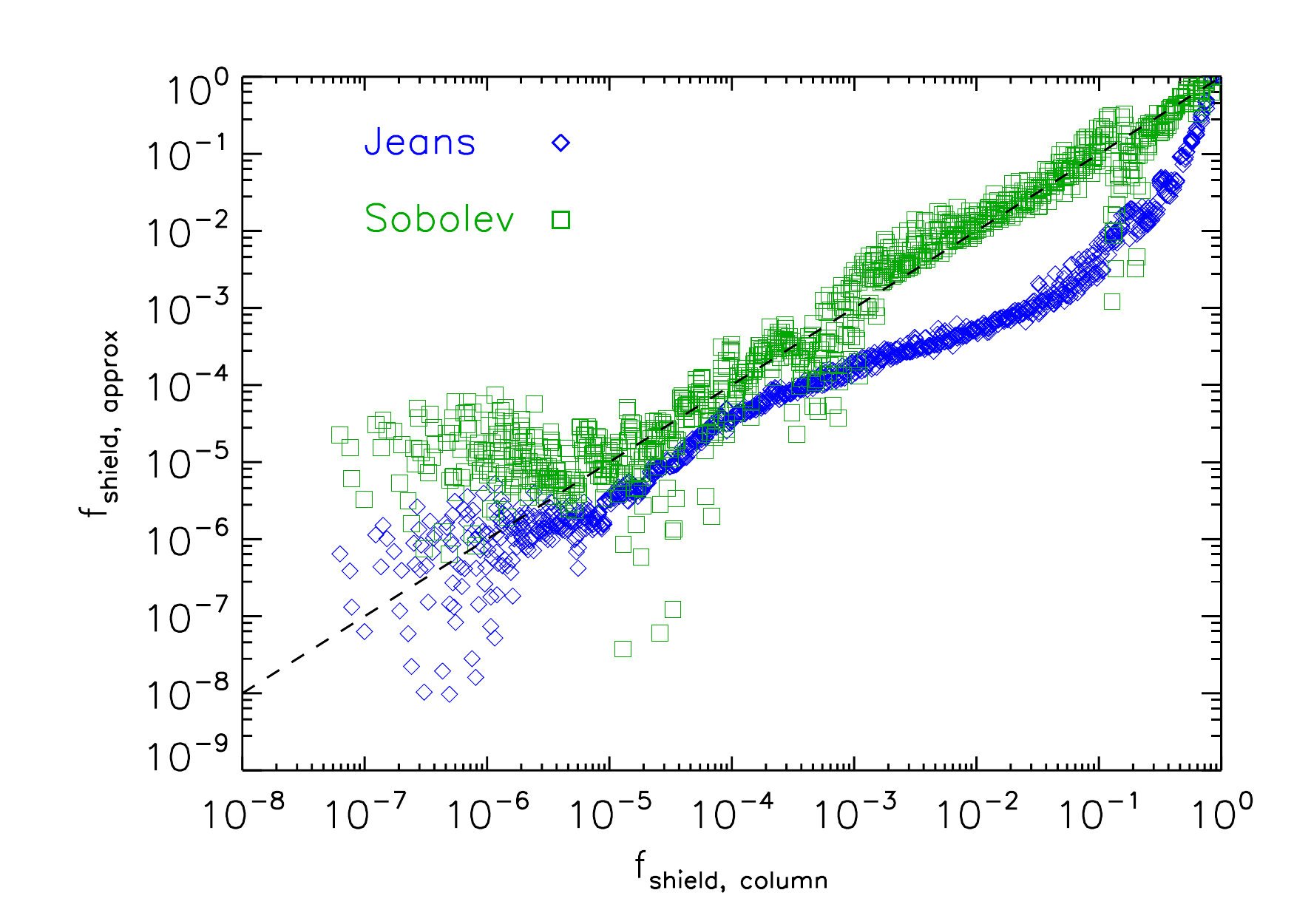}
\caption{Comparison of the $\htwo$ self-shielding factor, Equation (\ref{f_shield_h2_WG}), between approaches where the $\htwo$ column density $\nhtwo$ is calculated via our column density method ($f_{\mathrm{shield, \,\,column}}$) and two common approximations ($f_{\rm{shield, \,\,approx}}$). In these approximations, $\nhtwo$ is calculated locally with a Sobolev (green squares) and a Jeans length approach (blue diamonds). The dashed line represents a one-to-one mapping. The Sobolev approximation for $\nhtwo$ produces an $\htwo$ self-shielding factor in close agreement with our method, except for a few cases of disagreement, likely due to small velocity gradients. A hybrid of these two approaches may be useful in future work \citep[see][]{CLA11a}. }
\label{fig:fshield_compare}
\end{figure}

The photodissociation of $\htwo$ occurs through Solomon process \citep{SW67} in which a photon with an energy coinciding with either the Lyman or the Werner (LW) bands of $\htwo$ places the molecule in an excited electronic state. The subsequent radiative decay has a $\sim15\%$ chance of reaching the ground state continuum, which results in molecular dissociation. Accurately modeling this process requires detailed modeling of hundreds of LW lines which is computationally unfeasible in 3D hydrodynamic simulations. Fortunately, the photodissociation rate of molecular hydrogen can be expressed as \citep[e.g.,][]{AB97}
\begin{equation}
k_{\htwo} = 1.38\times10^{-12}\jlw\fshhtwo\,\, \mathrm{s}^{-1}\mbox{ ,}
\label{kh2}
\end{equation}
where the dimensionless factor
 $\fshhtwo\le1$ accounts for $\htwo$ self-shielding with $\fshhtwo=1$ corresponding to no shielding and $\fshhtwo=0$ to complete shielding. For a static, cold medium, $\fshhtwo$ can be written solely as a function of the $\htwo$ column density $N_{\htwo}$ \citep{DB96},
\begin{equation}
\fshhtwo = \mathrm{min}\left[1.0,\, \left(\frac{N_{\htwo}}{10^{14}\,\mathrm{cm}^{-2}}\right)^{-0.75}\right]\mbox{.}
\label{f_shield_h2_DB}
\end{equation}
To more accurately model a dynamic medium, \citet{DB96} also provided a fit for $\fshhtwo$ which takes thermal gas motion into account,
\begin{eqnarray}
 \label{f_shield_h2_WG}
 \fshhtwo & = &  \frac{0.965}{(1 + x/b_5)^{\alpha}}   \\
  & & {} +  \frac{0.035}{(1+x)^{0.5}} \times \mathrm{exp}[-8.5\times10^{-4}(1+x)^{0.5}]\mbox{ ,} \nonumber
\end{eqnarray}
where $x = N_{\htwo} / 5\times10^{14} \,\mathrm{cm}^{-2}$, $b_5 = b / 10^5\, \mathrm{cm}\,\mathrm{s}^{-1}$, $\alpha=2$, and $b=9.12\,\mathrm{km}\,\mathrm{s}^{-1}\, (T/10^4\,\kelvin)^{1/2}$ is the velocity spread parameter for $\htwo$ \citep[e.g.,][]{AS07}. Unless otherwise noted, we will use Equation (\ref{f_shield_h2_WG}) with $\alpha=1.1$, a modification suggested by \citet{WG11}, to calculate the self-shielding factor for $\htwo$ from LW radiation.

Hydrogen deuteride can also be destroyed by LW radiation at a rate similar to that given in Equation (\ref{kh2}), 
\begin{equation}
k_{\mathrm{HD}} = 1.5\times10^{-12}\jlw \fshhd \fshhdhtwo\,\, \mathrm{s}^{-1}\mbox{ ,}
\label{khd}
\end{equation}
which includes both a factor due to HD self-shielding and a factor accounting for the shielding of HD by $\htwo$. The HD self-shielding factor is equivalent to that in Equation (\ref{f_shield_h2_WG}) with the HD column density $N_{\mathrm{HD}}$ replacing $N_{\htwo}$. Due to slight energy differences between the $\htwo$ and HD LW line centres, the shielding of HD by $\htwo$ does not effectively occur until there is a large $\htwo$ column density, $N_{\htwo} \gtrsim 10^{20}\,\mathrm{cm}^{-2}$. \citet{WG111} provided a fit for the HD shielding factor due to $\htwo$,
\begin{equation}
\fshhdhtwo = \frac{1}{(1 + x)^{0.238}}\,\mathrm{exp}(-5.2\times10^{-3} \,x)\mbox{ ,}
\label{f_shield_hd}
\end{equation}
where $x = N_{\htwo} / 2.34\times10^{19} \,\mathrm{cm}^{-2}$. Both $\htwo$ and HD can also be shielded from LW radiation by neutral hydrogen, however we neglect this effect as self-shielding is always the dominant effect. In fact, HD photodissociation is never significant in determining the HD abundance. Instead, as we will argue, it is primarily the $\htwo$ photodissociation and self-shielding that determine both the $\htwo$ and HD abundances.

We compute the molecular column densities, $\nhtwo$ and $\nhd$, using an on-the-fly, non-local approach very similar to the `six-ray' approximation \citep[e.g.,][]{NL97,GM07,GLO10}. Specifically we compute the column density at each point in our box as
\begin{equation}
N_{\htwo}(\bmath{r}) = \mathrm{min}_j\{N_{\htwo,j}(\bmath{r}) \}\mbox{ ,}
\label{eq:nh2-1}
\end{equation}
where $j=\pm x, \pm y, \pm z$, and $N_{\htwo, j}$ is the column density from point $\bmath{r}$ to the edge of the box along direction $j$. The same approach is used for calculating the HD column density. The presence of bulk velocity gradients, which our approach does not account for, may act to decrease the importance of self-shielding by Doppler shifting absorption line centres \citep[e.g.,][]{DB96,GB01}, although \citet{WG11} argue that this effect is minimal in radially coherent gas flows.

While we self-consistently compute the column densities to each point in our simulation, a commonly used \citep[e.g.,][]{BL03,SHA10}, and less computationally expensive, approximation is $\nhtwo \approx n_{\htwo}\,L_{\mathrm{char}}$, where $L_{\mathrm{char}}$ is a local characteristic length scale and $n_{\htwo}$ is the number density of molecular hydrogen. In Figure \ref{fig:fshield_compare}, we compare $\fshhtwo$ computed using $\nhtwo$ calculated with our six-ray approach to $\fshhtwo$ computed with $\nhtwo$ based on two local characteristic lengths $L_{\mathrm{char}}$ at the time when the gas density first reaches $10^8\,\cc$. The first is the Jeans length, $L_{\mathrm{char}} = \lj$, which is motivated by the assumption that the majority of shielding occurs in a region of size similar to the local Jeans length. The second approximation is a close analogue to the Sobolev length \citep{Y06,CLA11a}, $L_{\mathrm{char}} = \cs / |\bmath\nabla \cdot \bmath{v}|$, which, assuming a constant velocity gradient, is the distance at which absorption line centres are Doppler-shifted by approximately one thermal line width. As is evident in Figure \ref{fig:fshield_compare}, both approximations agree reasonably well with the more accurate, non-local approach, with the Sobolev approach yielding slightly better overall agreement. While this agreement is likely coincidental given the disparate physics involved in all three approaches, it is still reassuring that they are all of similar magnitude. At least in this regime, a combination of the two local approaches could be useful in future work \citep[e.g.,][]{CLA11a}.

%
%

\section{Results}
\label{sec:results}

\begin{figure}
\includegraphics[scale=0.5, clip, trim = 20 5 5 5 ]{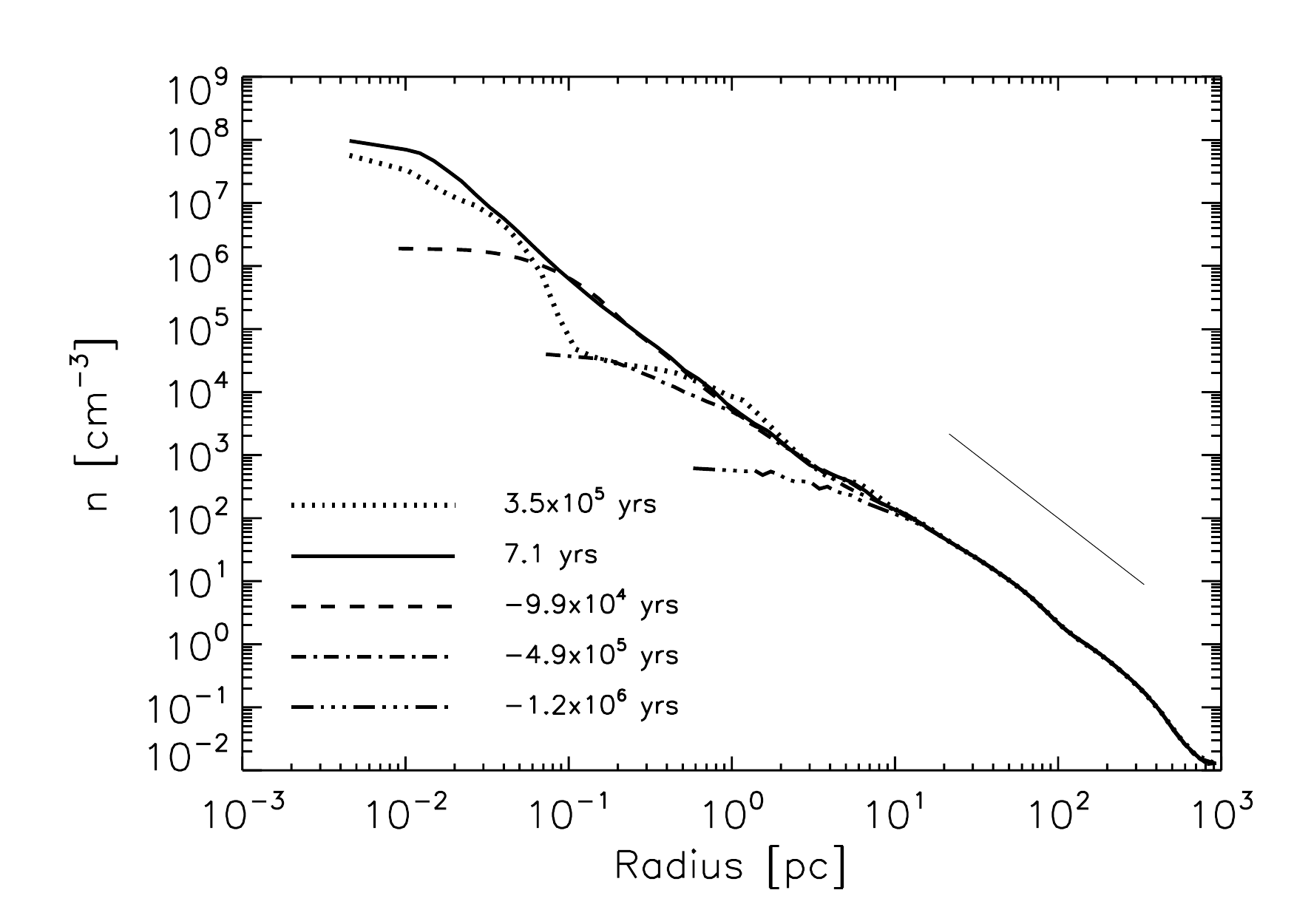}
\caption{Radial profiles of average gas density centered around the point of maximum density with time measured from the point of sink particle formation which occurs shortly after the gas first reaches $n=10^8\,\cc$. The gas density always maintains a roughly $\rho\propto r^{-2}$ density profile (shown by the straight line) with departures from the formation of the core and the gravitational influence of the sink particle at the final time shown. }
\label{fig:radial_dens_profiles}
\end{figure}

The LW radiation field of intensity $J_{21}=100$ prevents the $\htwo$ abundance from reaching the level that would permit efficient gas cooling in any halo not capable of atomic line cooling.\footnote[1]{If we included no radiation background, the first minihalo in our box would have collapsed at $z\sim19$; see \citet{RIT12}, who used identical cosmological initial conditions. } Hydrostatic collapse still occurs in these haloes, but only along an adiabat. At a redshift $z\sim13$, our target halo can be classified as an atomic cooling halo, with an average gas temperature of $8000\,\kelvin$ and most of gas within the virial radius lying between $2000\,\kelvin<T<1.2\times10^4\,\kelvin$. With gas now able to radiate its internal energy, collapse proceeds isothermally at $T\approx8000\,\kelvin$ resulting in an $n\propto r^{-2}$ density profile within the virial radius --- see Figure \ref{fig:radial_dens_profiles}.

When the maximum gas density near the halo centre reaches $\sim200\,\cc$ at $z=12.1$, the virial mass of the surrounding halo is $3\times10^7\,\solarmass$ with a virial radius $\approx750\,\pc$, just nominally fulfilling the standard analytic atomic cooling criterion, Equation (\ref{eq:T_vir}). At this point, the $\htwo$ cooling rate first exceeds the rate of adiabatic heating, resulting in rapid gas cooling to $T\sim400\,\kelvin$. 
 It should be noted that $\htwo$ self-shielding is not the determining factor in setting the density at which $\htwo$ cooling becomes effective. Once that happens, however, the self-shielding factor rapidly drops below unity indicating strong shielding. Indeed, the inclusion of self-shielding is essential for the cooling instability to continue and for the $\htwo$ abundance to eventually reach the asymptotic abundance of $\sim10^{-3}$ \citep{OH02}. 
 
In a suite of cosmological simulation exploring the effect of LW radiation on the fraction of cool gas in haloes, \citet{MAC01} determined the threshold halo virial mass for metal-free gaseous collapse as a function of radiation intensity to be
\begin{equation}
M_{\mathrm{TH}} = \{1.25\times10^5 + 2.9\times10^6[\jlw]^{0.47}\}\,\solarmass \mbox{ ,}
\label{eq:threshold_halo_mass}
\end{equation}
where we have converted their $F_{\mathrm{LW}}$ into $\jlw$. Extrapolating beyond the intensity range, $\jlw < 0.080$, explored by \citet{MAC01}, Equation (\ref{eq:threshold_halo_mass}) still yields an accurate prediction for the halo mass $M\approx3\times10^7\,\solarmass$ where self-shielding and collapse first occur in the simulation here. Since \citet{MAC01} did not run simulations with stronger radiation fields, their simulations did not produce any haloes in which atomic line cooling was effective. Indeed, once an atomic cooling halo forms, $\lya$ cooling permits rapid isothermal collapse provided that the halo mass exceeds a minimum that depends very weakly on the LW radiation intensity.

\begin{figure}
\includegraphics[scale=0.52, clip, trim = 30 10 0 0  ]{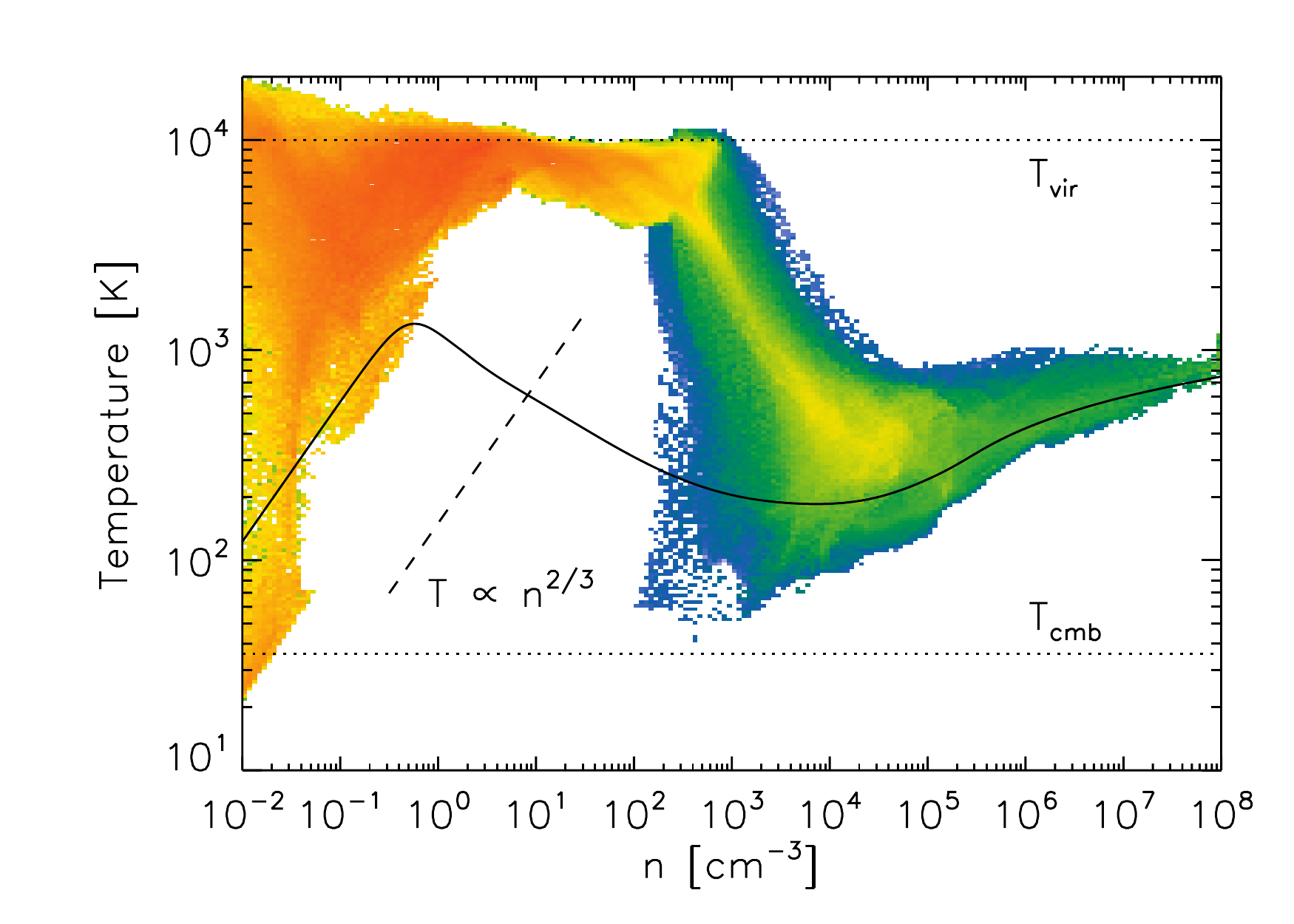}
\caption{Density-temperature phase plot at the time of sink particle formation at $z=12.1$ showing gas within $2R_{\mathrm{vir}}$ of the halo. The colour corresponds to the mass contained in the phase-space region. The solid black curve is a one-zone calculation of a thermodynamic evolution of metal-free gas undergoing free-fall collapse with no radiation background (Pop III.1 track). After the gas in the simulation is able to cool via $\htwo$ cooling starting around $n\sim2\times10^2\,\cc$, most of the cooled gas reaches minimum temperature at $T\sim400\,\kelvin$ and $n\sim10^4\,\cc$. The gas that continues to collapse converges towards the standard Pop III.1 molecular track. Adiabatic gas cooling due to expansion in a turbulent medium is apparent around $n\sim10^3\,\cc$, permitting some gas to reach temperatures $T<100\,\kelvin$.}
\label{fig:dens_temp_full}
\end{figure}

We will denote the density at which the $\htwo$ cooling rate becomes larger than the rate of adiabatic heating and the gas evolution leaves the atomic isothermal track with $\ncool$. We show a representative density-temperature phase diagram in Figure \ref{fig:dens_temp_full} which shows this density is $\ncool\approx200\,\cc$ and this density is found within a radius of $\sim10\,\pc$. This is a larger density than $\ncool\approx10\,\cc$ found by \citet{SHA10} in a similar simulation with an identical radiation background. We attribute this discrepancy to our different treatments of $\htwo$ self-shielding, in particular the method for computing $\nhtwo$. \citet{SHA10} utilized a local approximation $\nhtwo = n_{\htwo} \, \lj$, while here we computed $\nhtwo$ more accurately as described in Section \ref{sec:lw_radiation}. To further understand the source of this discrepancy, we can equate the $\htwo$ gas cooling time $\tcoolhtwo$ and the free-fall time $\tff$ to obtain a rough estimate of $\ncool$,
\begin{equation}
 \ncool  \approx  180\,\cc\,\left(\frac{\abundhtwo}{3\times10^{-7}}\right)^{-2} \mbox{ ,}
\label{eq:ncool1}
\end{equation}
where $\abundhtwo$ is the $\htwo$ abundance at $\ncool$, $\tff = [3\pi / (32G\rho)]^{1/2}$ is the free-fall time, $\tcoolhtwo= 3n\kb T/2\Lambda_{\htwo}$ is the $\htwo$ cooling time, and $\Lambda_{\htwo}$ is the volumetric $\htwo$ cooling rate \citep[e.g.,][]{GP98} which we evaluate at $8000\,\kelvin$. Equation (\ref{eq:ncool1}) is simply $\ncool$ in terms of the required $\htwo$ abundance for the molecular cooling rate to exceed the rate of adiabatic heating, where $3\times10^{-7}$ is the $\htwo$ abundance at which the transition to efficient molecular cooling takes place in our simulation. We can simplify this further by reasonably assuming that the $\htwo$ abundance is determined by its photoequilibrium value,
\begin{equation}
 x_{\htwo,\mathrm{photo}} = \frac{\kfhminus\,\abunde}{k_{\htwo}}n\mbox{ ,}
\label{eq:h2photoabund}
\end{equation}
where $\kfhminus$ is the formation rate of $\mathrm{H}^-$, $k_{\htwo}$ is the photodissociation rate of $\htwo$ given by Equation (\ref{kh2}), and $\mathrm{H}^-$ photodestruction is considered negligible. Inserting Equation (\ref{eq:h2photoabund}) into Equation (\ref{eq:ncool1}) and assuming $\fshhtwo=1$ results in
\begin{equation}
 \ncool \approx 210\,\cc\,\left(\frac{J_{21}}{100}\right)^{2/3}\left(\frac{\abunde}{5\times10^{-5}}\right)^{-2/3} \mbox{ ,}
\label{eq:ncool2}
\end{equation}
where $J_{21}$ is the unattenuated radiation intensity and $\abunde$ is the free electron abundance at $\ncool$. Equation (\ref{eq:ncool2}) is in excellent agreement with the density at which $\htwo$ cooling becomes effective in the simulation, suggesting that self-shielding is not necessary for the onset of $\htwo$ cooling. Allowing for self-shielding, $\fshhtwo\le1$, and adopting the prescription of \citet{SHA10} with Equation (\ref{f_shield_h2_DB}) and $\nhtwo = n_{\htwo} \, \lj$ gives us
\begin{equation}
\ncool \approx 10\,\cc\,\left(\frac{J_{21}}{100}\right)^{2/3}\left(\frac{\abunde}{10^{-4}}\right)^{-2/3} \mbox{ ,}
\label{eq:ncool3}
\end{equation}
where we have now normalized to a slightly higher value of the electron abundance found in the simulation at lower densities.
Equations (\ref{eq:ncool2}) and (\ref{eq:ncool3}) separately agree reasonably well with our simulation and with that of \citet{SHA10}, respectively, suggesting that the $\htwo$ column density computed with the local Jeans length is an overestimate compared to that computed with the more detailed radiative transfer described in Section \ref{sec:lw_radiation}. As we shall argue in Section \ref{sec:direct_observations}, this may have an impact on the star formation efficiencies and observability of these systems.

\begin{figure}
\includegraphics[scale=0.45, clip, trim = 5 15 30 5  ]{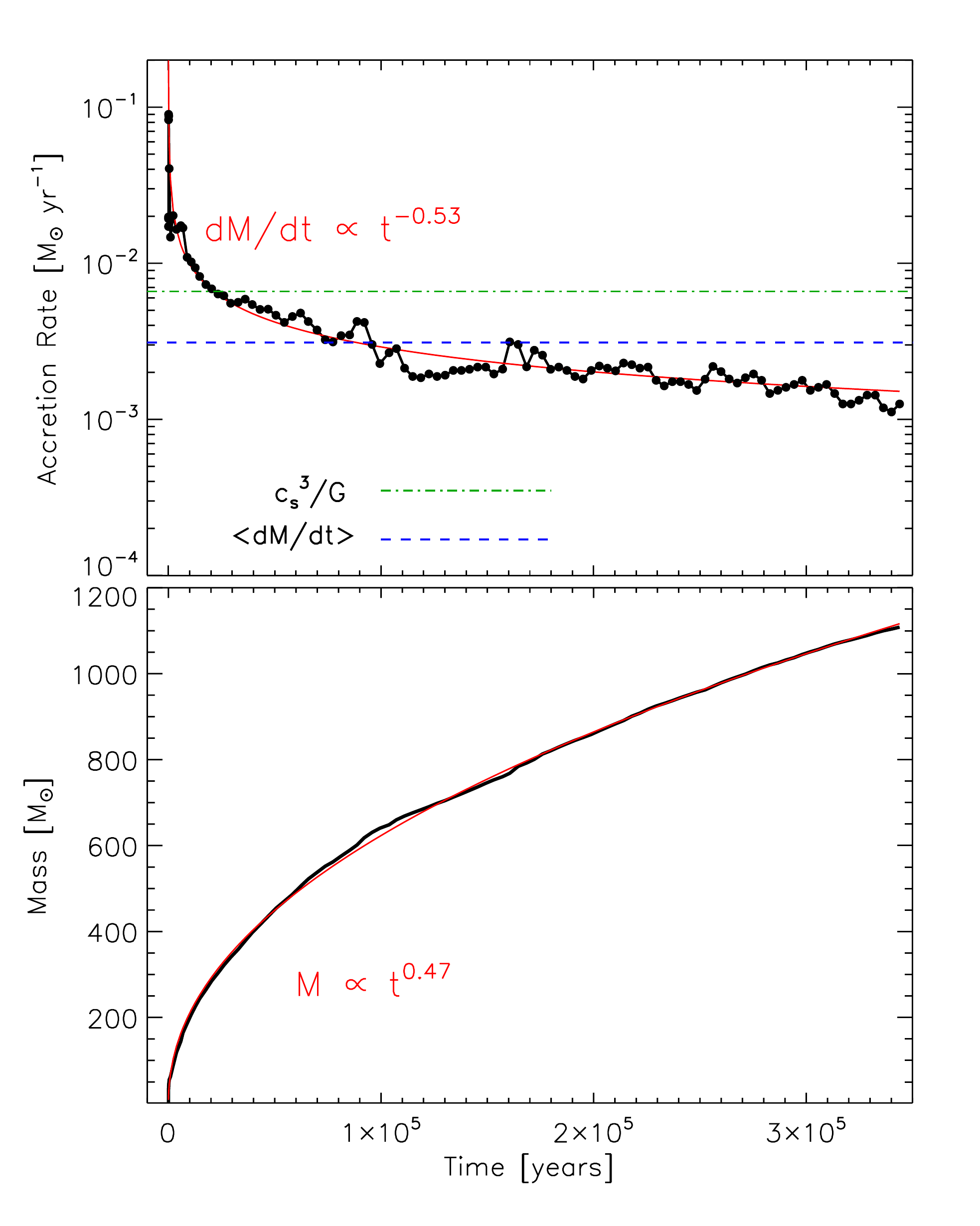}
\caption{Accretion rate (top panel) and mass (bottom panel) of the sink particle as a function of time. Power law fits to the accretion rate and mass in the first $10^5$ years are shown in red. The time-averaged accretion rate (blue dashed line) of the sink particle is $3\times10^{-3}\solarmass\,\yr^{-1}$, which we take to be an upper limit on the star formation rate that occurred in this time period. The late time accretion rate approaches $10^{-3}\solarmass\,\yr^{-1}$. We also show the characteristic accretion rate $\cs^3/G$ (green dot-dashed line) with the adiabatic sound speed $\cs$ evaluated at $T=800\,\kelvin$.}
\label{fig:sink_mass_accretion_rate}
\end{figure}

\begin{figure*}
\begin{center}
\includegraphics[scale=0.28, clip, trim = 0 0 0 0  ]{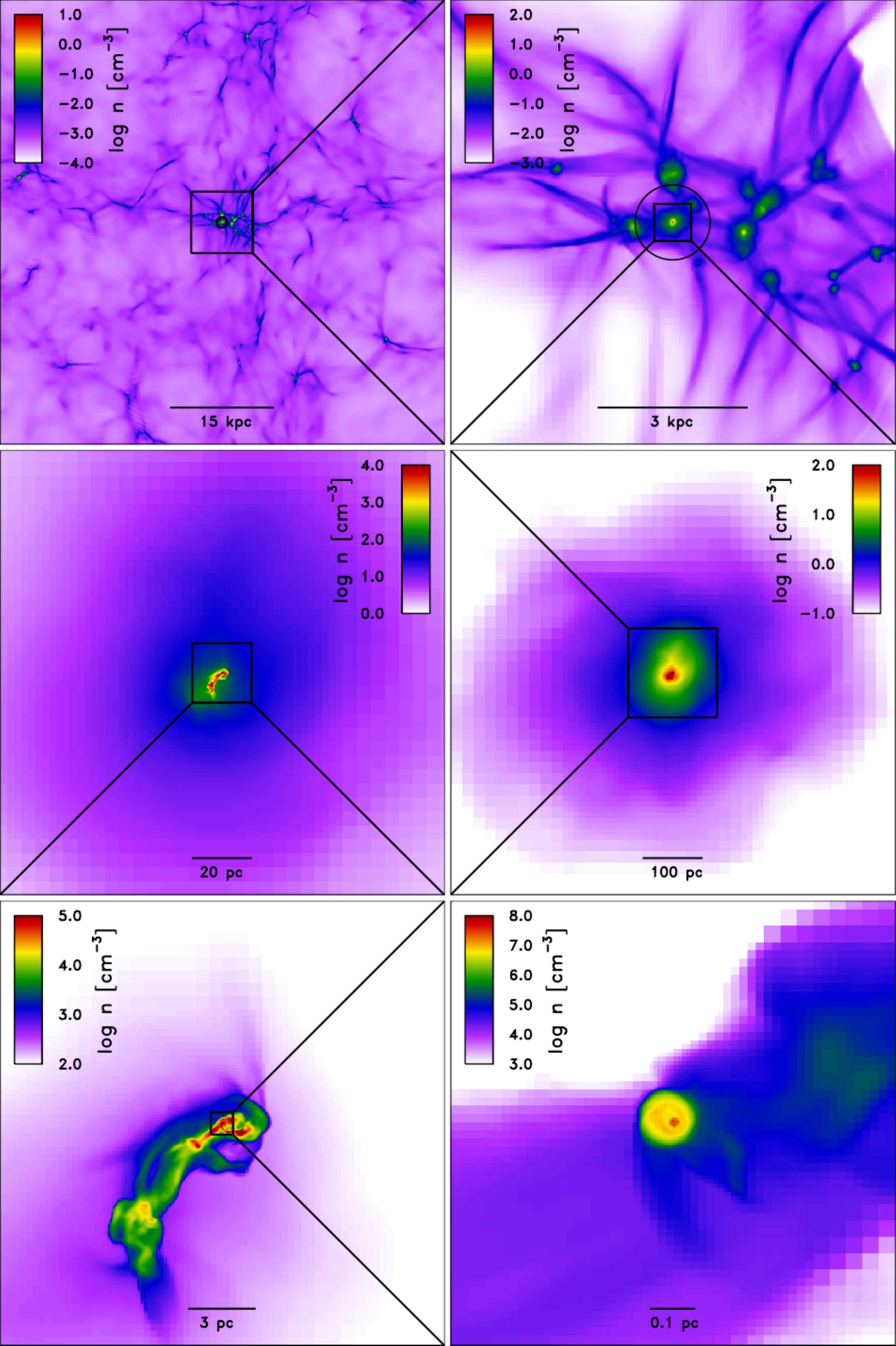}
\caption{Mass-weighted line-of-sight gas density projections $3.5\times10^5$ years after the sink particle formation ($z=12.1$). The six panels show progressively smaller fields of view. The black dot marks the location of the sink particle while the black circle in the top panels is the virial radius ($\sim750\,\pc$) of the target halo. The spatial scale, in physical units, is shown at the bottom of each panel. The upper-left panel shows neighboring haloes and the clustered cosmological environment where our target halo formed. Self-shielding, cold gas is seen in the bottom-left panel and the filamentary, irregular nature of this gas is apparent. The high density disc which forms around the sink particle is visible in the bottom-right panel, approximately face-on.}
\label{fig:zoomed_dens_projections}
\end{center}
\end{figure*}

The onset of effective $\htwo$ cooling at $\ncool$ leads to the rapid emergence of a cold $<10^3\,\kelvin$, dense $n\sim10^4\,\cc$ core, similar to that forming in the process of Pop III.1 star formation \citep[e.g.,][]{BCL02}. While we have argued that self-shielding is not important in determining $\ncool$, the $\htwo$ self-shielding factor does begin to drop below unity as the gas cools, reaching as low as $\fshhtwo\approx10^{-6}$ at the highest densities. Proper treatment of self-shielding is therefore essential for accurately computing the chemical state of high density gas. To discriminate cold, molecule-rich, self-shielding gas from the warmer outer halo, we will generally utilize a simple criterion where we select the cells that have an $\htwo$ shielding factor $\fshhtwo <10^{-2}$. Some of our results may be sensitive to how we decide which gas belongs to the self-shielding core, although we have verified that alternative criteria, such as only selecting gas with temperatures $<2\times10^3\,\kelvin$ or densities $>10^3\,\cc$, would yield similar conclusions.

\begin{figure}
\includegraphics[scale=0.55, clip, trim = 35 10 20 20  ]{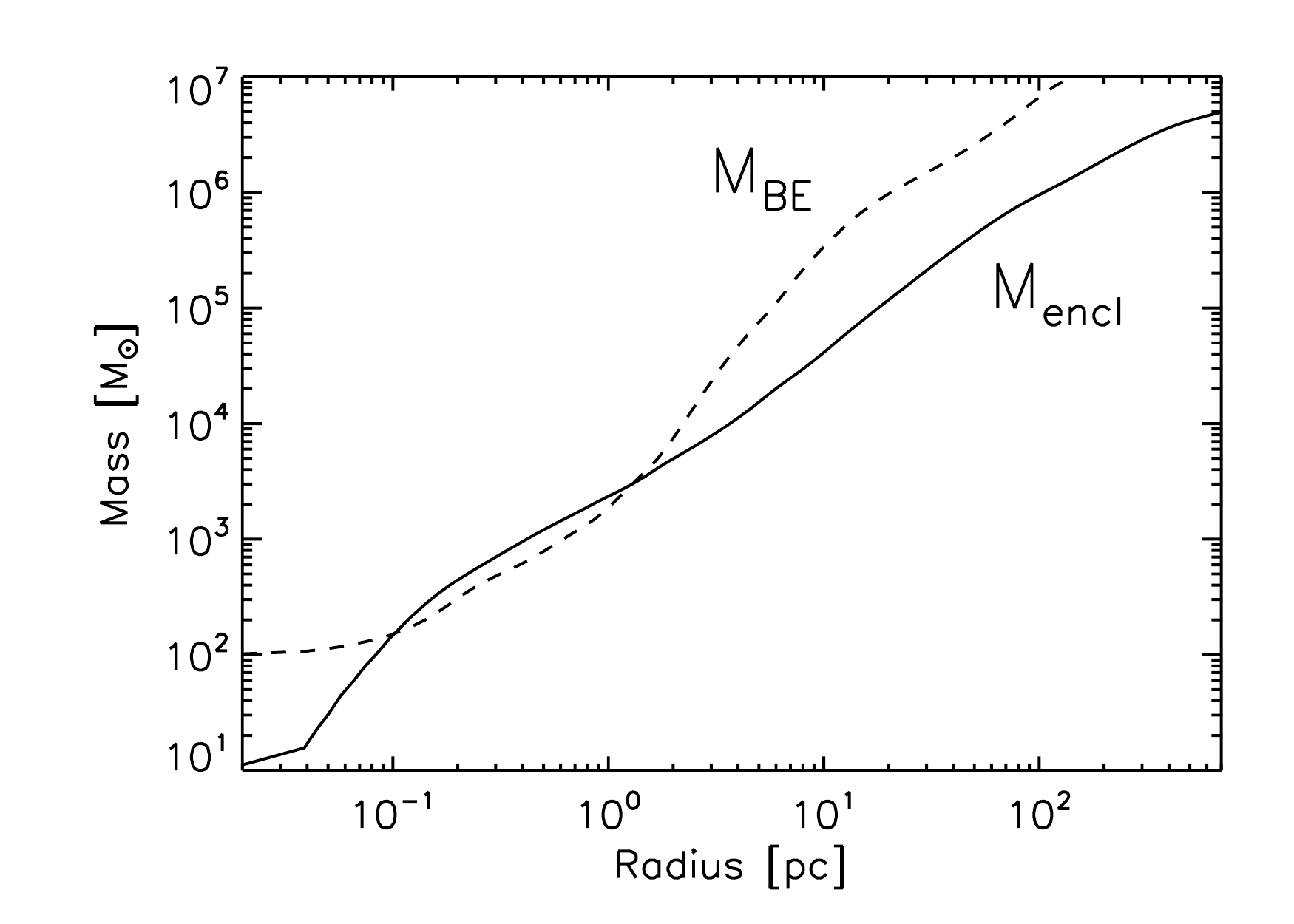}
\caption{Enclosed gas mass (solid line) and Bonnor-Ebert mass (dashed line) as a function of distance from the point of maximum density roughly $10^5$ years before sink particle formation. The maximum radius where the enclosed gas mass exceeds the Bonnor-Ebert mass provides a rough measure of the mass of the first gravitationally unstable clump. We see the onset of gravitationally instability occurring inside a radius of $\sim1\,\pc$ when the enclosed gas mass is $10^3\,\solarmass$.}
\label{fig:be_stability}
\end{figure}

After effective cooling begins when the gas density reaches $\ncool$, gas collapses at close to the free-fall rate and reaches $n=10^{8}\,\cc$ in $\sim3\,\mathrm{Myr}$. As the collapse progresses, we can estimate the mass of the first gravitationally unstable clump by comparing the enclosed gas mass to the Bonnor-Ebert mass \citep{EB55,BO56} at different radii. The Bonnor-Ebert mass can be written as

\begin{equation}
M_{\mathrm{BE}} = \frac{m_1 a_{T}^4}{P_0^{1/2}G^{3/2}} \mbox{ ,}
\label{eq:BE_mass}
\end{equation}
where $a_T = (\kb T / \mu \mh)^{1/2}$ is the isothermal sound speed, $P_0$ is the ambient pressure, and $m_1=1.18$ is the maximum dimensionless mass in the solution to the Lane-Emden equation \citep[e.g.,][]{SP05}. In Figure \ref{fig:be_stability} we show the enclosed gas mass (solid line) and the Bonnor-Ebert mass (dashed line) both as a function of distance from the point of maximum gas density, approximately $10^5$ years before sink particle formation. To evaluate the Bonnor-Ebert mass as a function of radius, $M_{\mathrm{BE}}(r)$, we take $P_0$ to be the pressure at radius $r$ and $a_T$ to be the mass-weighted isothermal sound speed interior to $r$. The enclosed gas mass first exceeds the Bonnor-Ebert mass at a radius of $\sim1\,\pc$, corresponding to a mass of $\sim10^3\,\solarmass$. We thus expect the first gravitationally unstable fragment to have roughly this mass, though there remains the possibility that additional fragmentation may occur on unresolved scales or at later times.  

 Sink particles are allowed to form at a density of $n=10^8\,\cc$ to follow the gas evolution for multiple free-fall times (see Section \ref{sec:sink_particles}). We run the simulation for $3.5\times10^5\,\yrs$ after the formation of the sink particle. At this point, the self-shielding core has an average density $\approx8\times10^3\,\cc$, total mass $\approx1.7\times10^4\,\solarmass$, mass-weighted root-mean-square (rms) velocity $\vrms\approx7.1\,\kms$, average temperature $\approx480\,\kelvin$, and rms Mach number $\mach = \vrms/\cs \approx 3.3$. Additionally, the sink particle has increased in mass to $\approx1100\,\solarmass$, which is $\approx6\%$ of the total mass of the self-shielding gas. No additional sink particles form, meaning that additional gravitational fragmentation, if any, would have been confined to occur on unresolved scales inside the sink particle.

In Figure \ref{fig:sink_mass_accretion_rate}, we plot the accretion rate and mass of the sink particle as a function of time. Initially, the sink particle accretion rate is $\dot{M}_{\mathrm{sink}}\sim0.1\,\msunperyr$ which drops to $0.01\,\msunperyr$ within $10^4\,\yrs$. The time averaged accretion rate over the whole simulation is $0.003\,\msunperyr$. Given the average density and mass of self-shielding gas at the end of the simulation, $n\approx10^4\,\cc$ and $\approx2\times10^{4}\,\solarmass$, respectively, the average rate of accretion onto the sink particle is roughly a factor of 10 lower than what would be expected if the gas had been undergoing free-fall collapse.  We discuss this inefficiency of gaseous collapse further in Section \ref{sec:fragmentation}.

\begin{figure}
\centering
\subfigure{
\includegraphics[scale=0.5, clip, trim = 30 10 0 20 ]{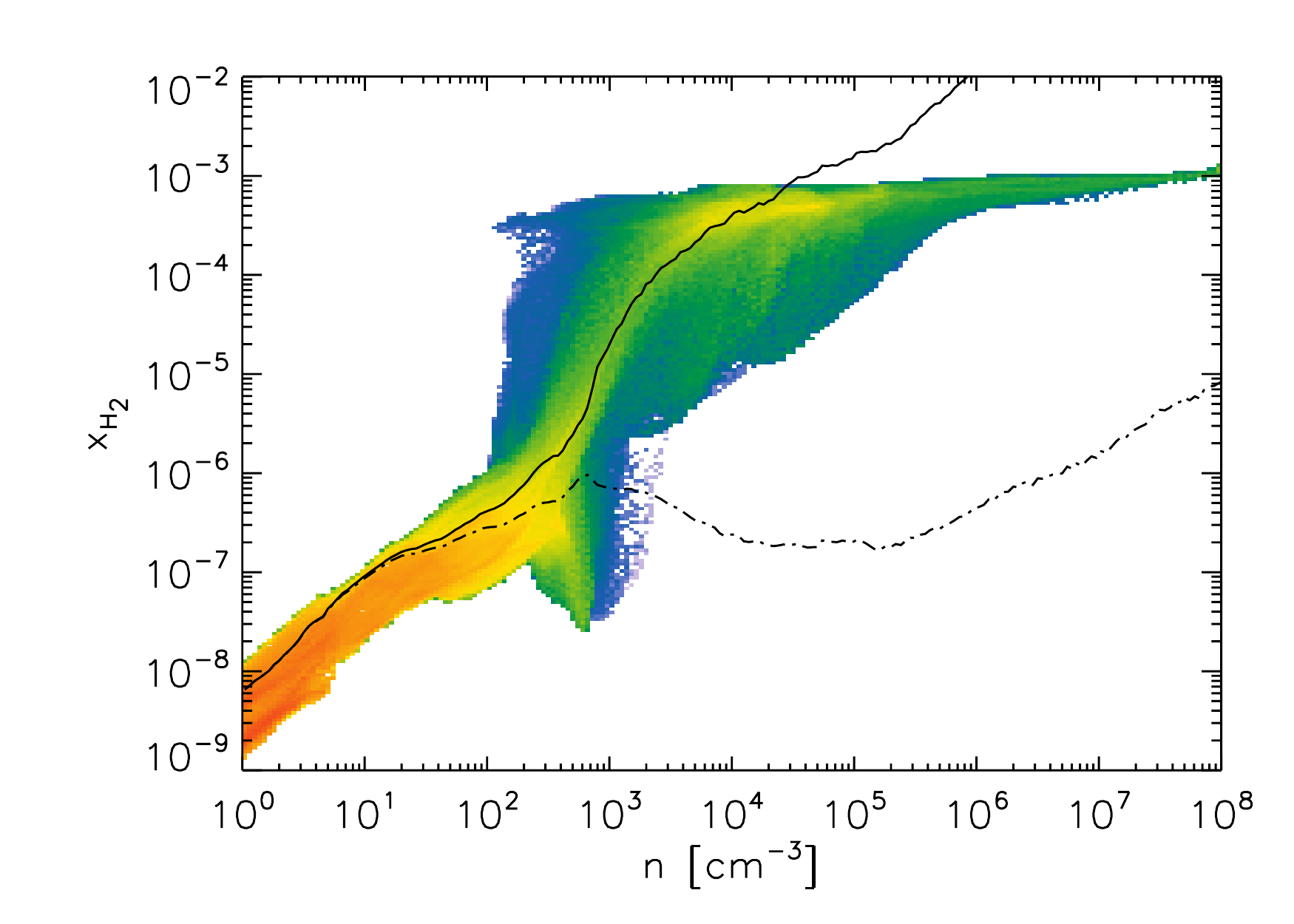}
\label{fig:subfig1}
}
\subfigure{
\includegraphics[scale=0.5, clip, trim = 30 10 0 20]{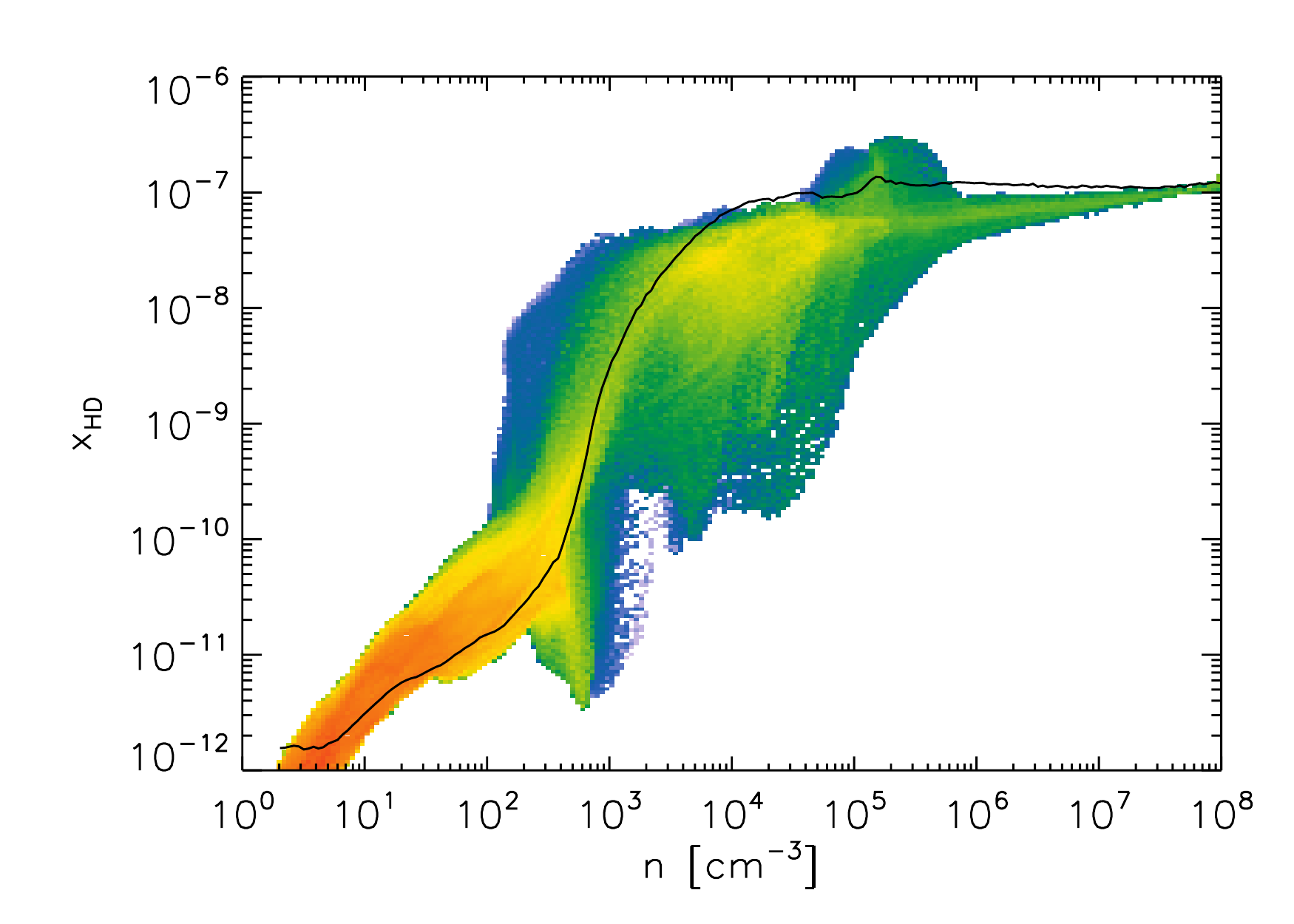}
\label{fig:subfig2}
}
\subfigure{
\includegraphics[scale=0.5, clip, trim = 30 10 0 20]{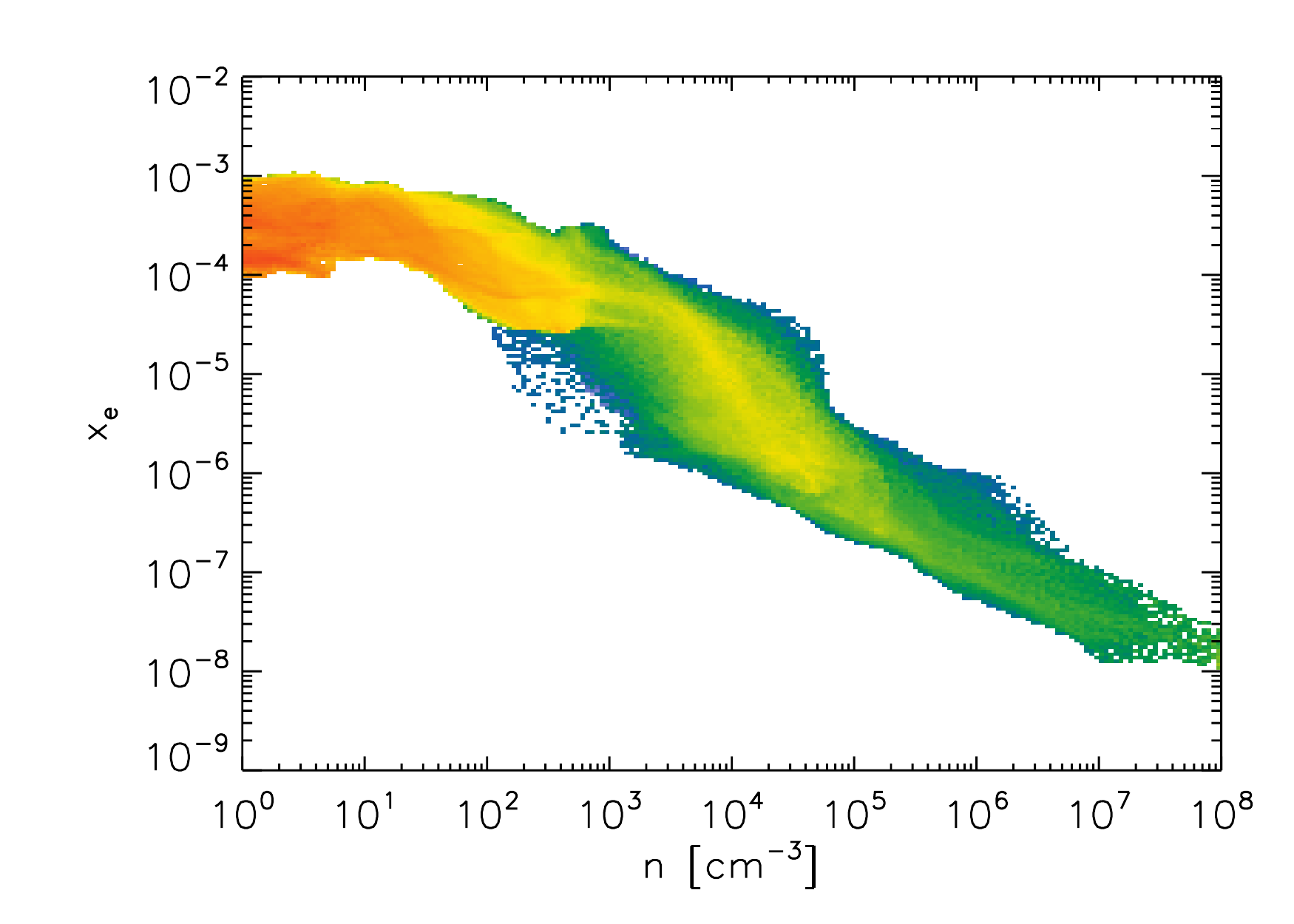}
\label{fig:subfig2}
}
\caption[]{ Abundance of $\htwo$ (top), HD (middle), and free electrons (bottom) as a function of density at the time when the sink particle formed. In the top panel, the dash-dotted line is the photo-equilibrium $\htwo$ abundance not considering self-shielding ($\fshhtwo = 1$) while the solid line includes self-shielding. As is shown, self-shielding is essential for $\htwo$ abundances larger than $\sim 10^{-6}$. In the middle panel, the black line is the equilibrium abundance of HD only assuming reactions involving $\htwo$ (see Equation \ref{eq:hd_eq}) and excluding HD photodissociation. The three equilibrium curves were computed with data extracted directly from the simulation and represent mass-weighted averages as a function of density.}
\label{fig:abundance_multiplot}
\end{figure}

 In Figure \ref{fig:zoomed_dens_projections} we show mass-weighted line-of-sight gas density projections on six different spatial scales $3.5\times10^5\,\yrs$ after the formation of the sink particle. The upper-right panel shows neighboring haloes and demonstrates the clustered cosmological environment where the target halo formed. The bottom-left panel displays the extent of self-shielding gas and its irregular, filamentary density structure. The self-shielding core is small, $\sim5-10\,\pc$, compared to the virial radius of the halo, $R_{\mathrm{vir}}\approx750\,\pc$. The bottom-right panel shows that a rotationally supported disc, $\sim0.1\,\pc$ in diameter, forms around the sink particle approximately $10^5$ years after the sink's formation. The disc is seen approximately face-on.

In Figure \ref{fig:abundance_multiplot} we show the abundances of $\htwo$, HD, and free electrons as a function of density at the time of sink particle formation. In gas with a density of $n=10^4\,\cc$, the $\htwo$ abundance reaches $\sim\few\times10^{-4}$ while the HD abundance is $\sim\few\times10^{-8}$. Both of these values approximately match equilibrium abundances, assuming that $\htwo$ is in direct photo-dissociation equilibrium with the self-shielding-attenuated LW radiation field and HD is close to equilibrium with $\htwo$ via the reaction in Equation (\ref{eq:hd_form}). We show the equilibrium abundance expectations for $\htwo$ and HD with solid black lines. Above this density, $\htwo$ approaches the non-equilibrium asymptotic abundance of $\abundhtwo\approx10^{-3}$ \citep[e.g.,][]{OH02} while HD remains in approximate equilibrium with $\htwo$. We observe only minor fractionation of HD over $\htwo$, which is to be expected given the relatively high temperature of the gas $\sim 400\,\kelvin$ compared to what is needed for the $\abundhd/\abundhtwo$ ratio to exceed the cosmological $\abundd/x_{\mathrm{H}}$ ratio. In fact, the HD abundance is generally just below its equilibrium value at densities $10^4\,\cc\lesssim n\lesssim10^8\,\cc $. The electron abundance shows the expected behavior --- when efficient $\htwo$ cooling begins at $n\sim10^2\,\cc$, the abundance is below the residual recombination value even though there is significant collisional ionization, $\abunde\sim10^{-2}$, at lower densities (not shown in Figure \ref{fig:abundance_multiplot}). We propose this as one explanation for the lack of significant HD cooling and discuss this further in Section \ref{sec:hd_cooling}.

\begin{figure*}
\begin{center}
\includegraphics[scale=0.6, clip, trim = 0 0 0 0  ]{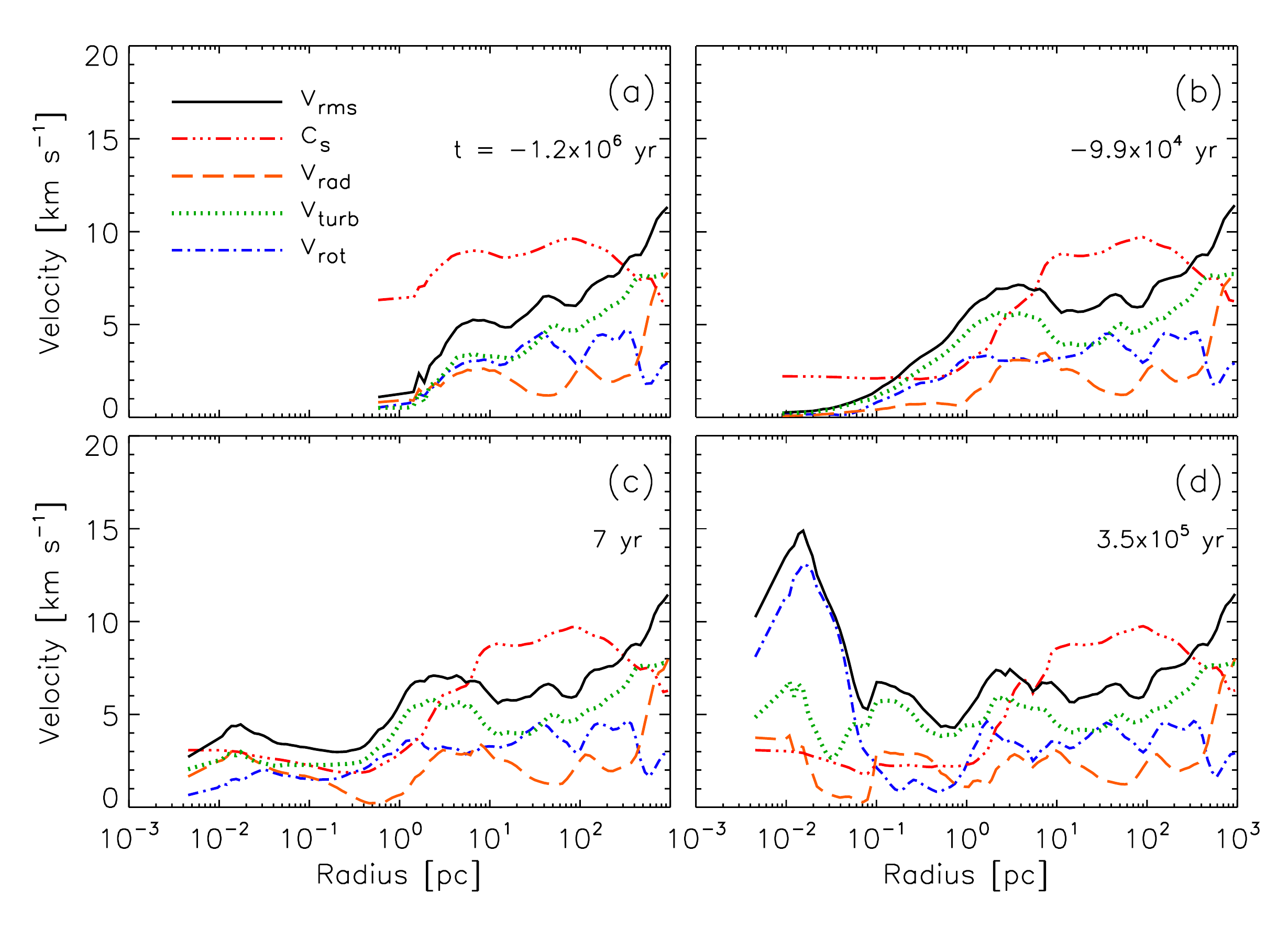}
\caption{Mass-weighted average gas velocities computed in annular shells. Panels (a) through (d) correspond to times $-1.2\times10^6, -9.9\times10^4, 7, \mathrm{ and } +3.5\times10^5\,\yrs$ from the formation of the sink particle. We show the rms velocity minus the centre of mass motion of the halo (black solid line), sound speed (red triple-dot dashed line), radial velocity (orange dashed line), rms turbulent velocity (green dotted line), and rms rotational velocity (blue dot-dashed line). See text for additional details. }
\label{fig:vel_rad_profs_evol}
\end{center}
\end{figure*}

%
%

\subsection{Kinematical State and Evolution}
\label{sec:velocity_profiles}

To gain further insight into the properties of the gas flow we proceed to analyze the gas kinematics at different stages of the collapse. Before the onset of effective $\htwo$ cooling, $\lya$ cooling acts as a thermostat that keeps gas inside the halo at a temperature $\sim8000\,\kelvin$. Towards the centre of the halo, isothermal collapse proceeds under quasi-hydrostatic conditions. However, with increasing density, the cooling time decreases faster than the free-fall time, eventually reaching a point at which $\htwo$ cooling is so rapid as to effectively remove pressure support and set the gas into free-fall collapse. This also marks the onset of self-shielding which allows the $\htwo$ abundance and cooling-rate to increase further. This thermal instability proceeds until the gas temperature is roughly $400\,\kelvin$. The collapse is also responsible for exciting some bulk turbulent motions and increasing the Mach number of the flow. At the time of sink particle formation, the self-shielding gas has an rms velocity $\vrms\approx7\,\kms$ corresponding to a Mach number $\mach\approx3$.

In Figure \ref{fig:vel_rad_profs_evol} we show different components of the gas velocity computed in annular shells at four times: (a) just before the onset of effective $\htwo$ cooling, (b) after the onset of $\htwo$ cooling but before the maximum gas density reached $10^8\,\cc$ and a sink particle formed, (c) at the moment of formation of the sink particle, and (d) at the end of our simulation, $3.5\times10^5$ years after sink formation. We consider five components of velocity, each computed via mass-weighting in annular shells centered on the sink particle location, except in panel (a) where it is centered on the point of maximum gas density:
\begin{enumerate} 
\item The rms velocity
\begin{equation}
\vrms = \left(\frac{\sum_{i}m_i|\bmath{v}_i|^2}{\sum_{i}m_i}\right)^{1/2}\mbox{ ,}
\label{eq:vrms}
\end{equation}
where the sum ranges over all cells with centres inside the annular shell, $m$ is the cell mass, and $\bmath{v}$ is the gas velocity minus the centre-of-mass motion of the particular annular shell. 

\item  The adiabatic sound speed $\cs = (\gamma\kb T / \mu \mh)^{1/2}$, where $T$ is the mass-weighted temperature inside the shell.
\item  The rms rotational velocity of the mean rotation inside the shell
\begin{equation}
\vrot = \left(\frac{ \sum_im_i| \bmath{v}_{\mathrm{rot},i} |^2   }{\sum_im_i}\right)^{1/2} \mbox{,}
\label{eq:vrot}
\end{equation}
where $\bmath{v}_{\mathrm{rot},i} = \bmath{r}_i\times\mathbf\Omega$ is the azimuthal rotational velocity of the mean flow in cell $i$, $\mathbf\Omega=\bmath{L} / I$ is the mean angular velocity inside the shell, $\bmath{L} =\sum_im_i\bmath{r}_i\times \bmath{v}_i$ is the total angular momentum inside the shell, and $I= \sum_im_i|\bmath{r}_i\times \bmath{\hat{L}}|^2$ is the moment of inertia of the shell in the direction defined by the total angular momentum.
\item  The radial velocity of the bulk motion inside the shell
\begin{equation}
\vrad = \frac{\sum_im_i  \bmath{\hat{r}}_i\cdot\bmath{v}_i }{\sum_im_i}\mbox{ .}
\label{eq:vrad}
\end{equation}
\item  The rms turbulent velocity relative to the mean flow, quantifying unordered gas motions
\begin{equation}
\vturb = \left(\frac{\sum_im_i|\bmath{v}_i - \bmath{v}_{\mathrm{rot},i} - \bmath{v}_{\mathrm{rad},i}|^2}{\sum_im_i}\right)^{1/2}   \mbox{,}
\label{eq:vturb}
\end{equation}
where $\bmath{v}_{\mathrm{rad},i} = \vrad\,\hat{\bmath{r}}_i$ is the radial velocity of cell $i$. 
\end{enumerate}

As shown in Figure \ref{fig:vel_rad_profs_evol}, before the onset of significant $\htwo$ cooling (panel a), typical rms gas velocities in the halo are $\sim10\,\kms$ near the virial radius and drop to $\sim5\,\kms$ in the inner $\sim10\,\pc$. The sound speed is greater than the rms velocity inside $300\,\pc$, somewhat smaller than half of the virial radius. Inside the virial radius bulk gas motions are dominated by turbulent and rotational motions, as radial infall motions greatly decrease after the virial shock. Gas kinematics inside the halo can be described as a turbulent, subsonic flow. Within $1$ Myr after the onset of $\htwo$ cooling (panel b), and shortly before sink particle formation, the mass of self-shielding gas is $\approx10^4\,\solarmass$. The temperature has dropped to $\approx400\,\kelvin$ in the inner $\sim1\,\pc$. The rms velocities have increased within $r\approx10\,\pc$ where the gas flow is transonic, $\mach\sim1$. At the time of sink particle formation (panel c), turbulent motions still dominate within $\sim10\,\pc$, although towards smaller radii the contributions of rotational and radial motions increase. The Mach number is on the order of $2$. Panel (d) shows the kinematical state of the gas $3.5\times10^5\,\yrs$ later at the end of the simulation. Turbulent motion is still the dominant form of gas motion, particularly in the range $0.1\,\pc\lesssim r \lesssim10\,\pc$. Within $r\sim0.1\,\pc$ there is kinematical evidence of a disc with ordered rotational velocities approaching $\sim15\,\kms$. The Mach number of the flow has also increased within the self-shielding core to $\mach\sim2-4$. We will proceed to analyze the impact of these supersonic gas motions on the collapse of gas in the following section.

%
%

\section{Supersonic Turbulence}
\label{sec:turbulence}

Supersonic turbulence is believed to play a key role in present-day star formation as it compresses gas to densities at which it can be susceptible to gravitational fragmentation and collapse. It additionally can act as a source of pressure support against gravitational collapse on larger scales \citep[e.g.,][]{MK04}. Minihaloes, the sites of Pop III.1 star formation, are not thought to contain fully developed supersonic turbulence \citep[e.g.,][]{Y06}. The virialization process, gravitational inflow, and larger virial velocities of atomic cooling haloes are believed to give rise to supersonically turbulent flows, likely influencing star formation \citep{WA07, GB08, PRI11a}. The turbulence which develops in the self-shielding core in our simulation is also generated by gravitational infall \citep[see][for a study of general properties of gravity-driven turbulence]{FED11}, but on a smaller spatial scale defined by thermal instability \citep[e.g.,][]{KN02}.

The classical theory of Kolmogorov \citep{KOL41,FRI95} describes turbulence as a chaotic fluid motion in which energy progressively cascades to smaller length scales through a series of eddies. These eddies eventually reach a small enough scale where viscous forces effectively dissipate the turbulence. Typical gas in the interstellar medium is highly compressible and compressible turbulence is known to result in a self-similar network of interacting shocks, large density contrasts, and a filamentary morphology \citep[e.g.,][]{KRI07}.

Many simulations have shown that driven, isothermal, supersonic turbulence results in a gas density probability distribution function (PDF) that is accurately described by a log-normal distribution \citep[e.g.,][]{VS94,PAS98,SC98,OS01}. This log-normal shape is understood to be the result of multiple, independent shocks altering the logarithmic gas density contrast in random walk fashion, thus driving the density PDF towards a log-normal form. The addition of self-gravity, magnetic fields, or a realistic equation of state can alter this general shape. 
We write the lognormal distribution as
\begin{equation}
p(s)\,ds = \frac{1}{(2\pi \sigma^2)^{1/2}} \,\mathrm{exp}\left[-\frac{1}{2}\left(\frac{s-\bar{s}}{\sigma}\right)^2\right]ds\mbox{ ,}
\label{eq:lognormal}
\end{equation}
where $p(s)$ is the probability that a parcel of gas has $s$ in the range $[s,s+ds]$, $s\equiv \mathrm{ln}(n/n_0)$ is the logarithmic density contrast, $\sigma$ is the standard deviation, and $n_0$ is the average density. Additionally, by considering the normalization of the log-normal distribution, it can be shown that $\bar{s} = -\sigma^2 / 2$. Unless otherwise noted, we consider \emph{volume}-weighted quantities when discussing density PDFs.

\begin{figure}
\includegraphics[scale=0.5, clip, trim = 20 0 0 0  ]{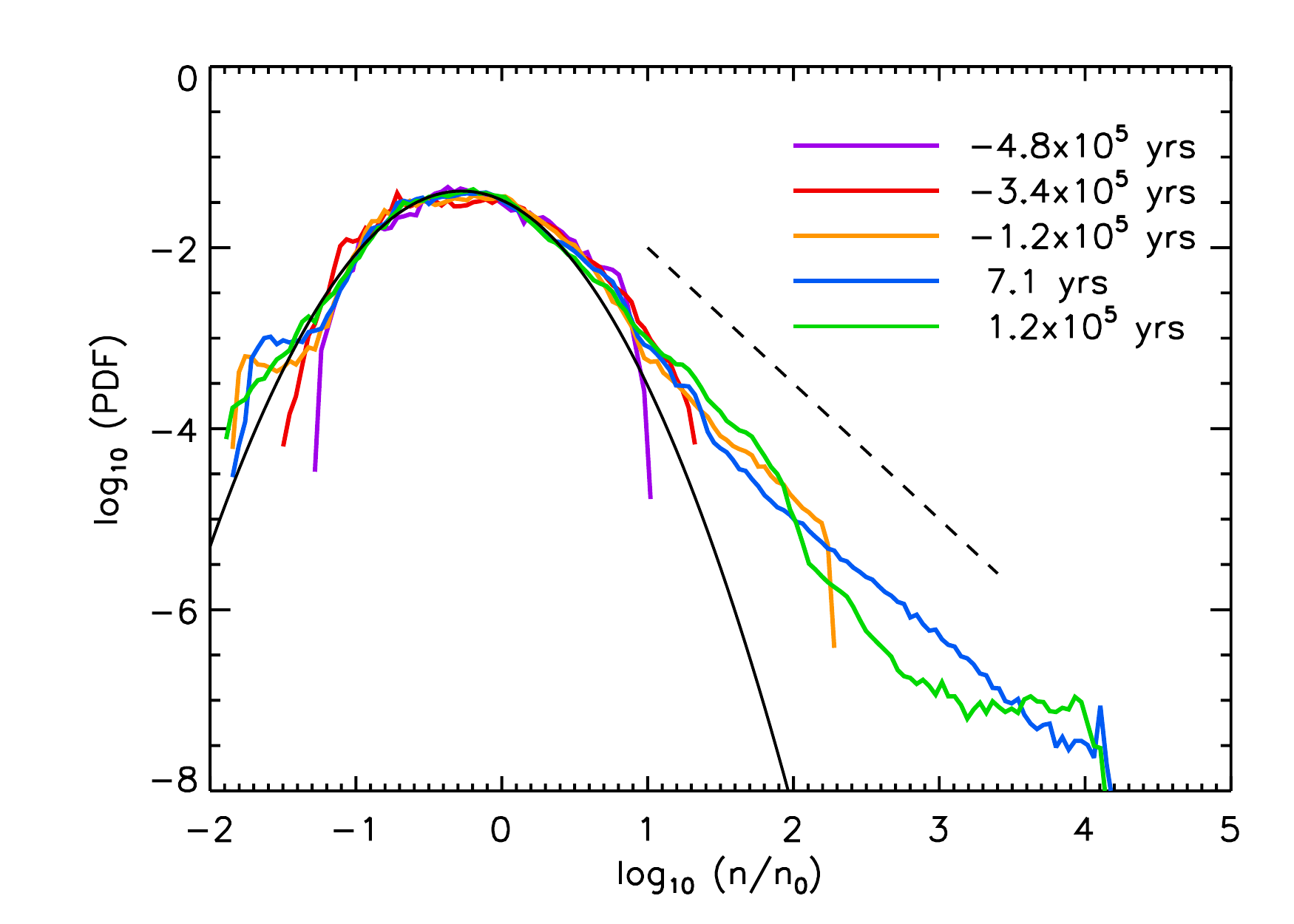}
\caption{Volume-weighted density PDFs of self-shielding gas ($ \fshhtwo < 10^{-2}$) at various times from the creation of the sink particle. We show a log-normal fit to the final density PDF (black curve) which has a standard deviation $\sigma=0.93$. The PDF populates higher density contrast as gas collapses and at high densities shows a marked departure from the initial log-normal like distribution which can be attributed to the self-gravity of the gas. The dashed black line indicates a power law of slope $-1.5$, the expected PDF slope of an $\rho\propto r^{-2}$ density profile. }
\label{fig:density_pdf_evol}
\end{figure}

In Figure  \ref{fig:density_pdf_evol} we show the volume-weighted density PDF of self-shielding gas, defined such that $\fshhtwo < 10^{-2}$, at five different times. As expected, the peaks of the density PDFs lie just below the average density as most of the volume is underdense, typical of compressive turbulence (the peak of a \emph{mass}-weighted density PDF would lie just above the average density). The black line is a log-normal fit to the final density PDF at $t=1.2\times10^5\,\yrs$, although such a fit to the PDF at any time would look approximately the same in this view. For this particular log-normal fit, $\sigma = 0.93$. 

As time progresses, the gas density PDF develops a power-law tail towards higher density contrasts. These power-law tails are understood as a consequence of the self-gravity of the gas and have been seen in both numerical simulations of turbulence \citep[e.g.,][]{SC98,KLE00,FED08a,COL11,KRI11,CK11} and observations of active star forming complexes \citep[e.g.,][]{KAI09,SCH12}.
It can be shown that the slope of the density PDF for a spherical $\rho\propto r^{-n}$ density profile is $-3/n$ \citep[e.g.,][]{KRI11,FED11}. Thus, for an isothermal, $n=2$, profile one would expect a power-law PDF with a slope of $-1.5$; this slope is shown in Figure \ref{fig:density_pdf_evol} as a dashed black line. The slope of the power-law tail just after sink particle formation matches this expectation quite well, while approximately $10^5\,\yrs$ later the gravity of the sink particle has altered the gas dynamics thus perturbing the shape of the PDF.

Numerous studies \citep[e.g.,][]{PAD97,PAS98,PRI11b,MOL12} have shown the rms Mach number $\mach$ and gas density PDF width $\sigma$ of a supersonically turbulent flow are related by
\begin{equation}
\sigma^2 = \mathrm{ln}(1 + b^2\mach^2) \mbox{ ,}
\label{eq:Mach_sigma}
\end{equation}
where $b$ is a constant, found numerically to lie between $\sim0.2-1.1$ \citep[see][]{FED08}. Since ordered rotational and radial motions do not directly produce density fluctuations, it is the turbulent Mach number, $\machturb = \vturb / \cs$ with $\vturb$ defined as in Equation (\ref{eq:vturb}), that should be present in Equation (\ref{eq:Mach_sigma}). Since the gas is dominated by turbulent motions after $\htwo$ cooling becomes effective (see Figure \ref{fig:vel_rad_profs_evol}), especially in low-temperature, self-shielding regions, $\vrms$ and $\vturb$ are very similar here. We can compare the parameters of the density PDF of the self-shielding gas in the simulation with expectations derived from the independently measured turbulent Mach number of the gas flow. Roughly $10^5\,\yrs$ after sink particle formation, the width of the density PDF is $\sigma=0.93$, corresponding to the green curve in Figure \ref{fig:density_pdf_evol}, and the turbulent Mach number in the self-shielding core is $\machturb\approx2.7$. Equation (\ref{eq:Mach_sigma}) would then require $b\approx0.43$, in excellent agreement with previous numerical experiments of idealized turbulence in which turbulence is driven by a combination of solenoidal (divergence-free) and compressive (curl-free) modes, i.e., mixed-mode forcing \citep[][]{FED10a}. This seems to suggest that the gravitational infall which is driving supersonic turbulent motions in the cold, self-shielding core, is exciting a combination of solenoidal and compressive modes, and that turbulence has modified the density structure of the gas in close accord with theoretical expectations.

We stress that log-normal density PDFs have been found in extremely idealized simulations where supersonic turbulence is driven by random forcing. In the simulation presented here, however, the turbulence is driven by gravitational infall and thermal instability. Thus, it is interesting, though not surprising, that the parameters of the gas density PDF in our cosmological simulation so closely match the idealized, theoretical expectations.

%
%

\section{Gas Fragmentation and Star Formation Rate}
\label{sec:fragmentation}

In this section we discuss trends towards gravitational fragmentation in our simulation. We also attempt to predict properties of the expected sub-grid star formation in the sink particle, including whether fragmentation on scales smaller than the sink particle accretion radius is expected. Our analysis is the first step towards learning about the mass spectrum of the stellar objects produced and the overall efficiency in which this system converts gas into stars.

\subsection{Star Formation Rate and Global Fragmentation}
\label{sec:star_formation_rate}

For $3\times10^5$ years after sink particle creation, the extent of our simulation, no additional sink particles form. At the end of the simulation the lone sink particle has a mass of $\approx1100\,\solarmass$, which is $\sim6\%$ of the mass of the self-shielding, cold gas. To understand this lack of fragmentation and gain insight into the expected rate of star formation, we define the dimensionless star formation rate per free-fall time as
\begin{equation}
\sfrff = \frac{\dot{M}_{\mathrm{sink}}}{(M/\tff)}  \mbox{ ,}
\label{eq:sfe}
\end{equation}
where $M$ and $\tff$ are the total mass and characteristic free-fall time of the self-shielding region, and $\dot{M}_{\mathrm{sink}}$ is the sink particle accretion rate (see Figure \ref{fig:sink_mass_accretion_rate}). $\sfrff$ is a measure of the actual rate of star formation compared to the rate if gas is collapsing on the free-fall timescale. Even though $\dot{M}_{\mathrm{sink}}$ is the sink accretion rate, not the star formation rate, it can be interpreted as the rate at which gas collapses to high densities where it can fuel star formation. Given that we do not attempt to model any feedback effects from the sink particles, such as protostellar outflows and radiation, $\dot{M}_{\mathrm{sink}}$ is a firm upper limit to the actual star formation rate. 

Taking fiducial values for the self-shielding core, namely the average sink particle accretion rate, characteristic self-shielding cloud density, and its mass Equation (\ref{eq:sfe}) becomes
\begin{eqnarray}
\label{eq:sfe1}
 \sfrff  & \approx & 0.1 \,\left(\frac{\dot{M}_{\mathrm{sink}}}{3\times10^{-3}\,\solarmass\, \yr^{-1}}\right) \nonumber  \\
  & & {} \times \left(\frac{\tff}{10^6\,\yr}\right)\left(\frac{M}{10^4\,\solarmass}\right)^{-1} \mbox{ ,}  
\end{eqnarray}
which suggests that star formation in our system is at least $10$ times slower than it would be if all the gas in the self-shielding core gas had been collapsing at the free-fall rate. We will argue this retardation stems from the supersonic turbulence and a centrifugally supported disc around the sink particle.

One way to understand how supersonic turbulence suppresses the rate of gaseous collapse is by considering it an additional source of pressure support. The Jeans mass, $\mj$, scales with sound speed and density like $\mj \propto \rho^{-1/2}\cseff^3$. The presence of supersonic turbulent motions enhances the effective sound speed, $\cseff^2 =  \cs^2 + \vrms^2/3$. Additionally, strong isothermal shocks impart density fluctuations scaling as $\rho\propto\mach^2 \propto \vrms^2$. These two effects lead to a scaling of $\mj\propto\vrms^2$ in supersonic gas, implying turbulence does indeed impede collapse by increasing the effective Jeans mass \citep[e.g.,][]{CHAN1951,BON87,MK04}. It should be noted, however, that this analysis assumes that the highly anisotropic turbulent velocity field is isotropic; thus treating supersonic turbulence as an additional source of pressure is not necessarily justified.

It has long been known that the global $\sfrff$ in the Galactic giant molecular clouds (GMCs) is on the order of $\sim 0.01$ \citep[e.g.,][]{ZUC74}. This low rate of star formation extends to more compact star-forming systems as well, such as infrared dark clouds (IRDCs) and higher density clumps embedded within \citep[see][and references therein]{KT07}. Equation (\ref{eq:sfe1}) suggests a global $\sfrff$ of $\sim0.1$ for self-shielding, cold gas seen in our simulation, ten times that of the Galactic norm. Perhaps this is not surprising, though, given we neither include any sort of stellar feedback nor magnetic fields, both of which would likely reduce the star formation rate. Additionally, the rms Mach number in the self-shielding gas here, $\mach\sim3$, is much less than typical Mach numbers in Galactic molecular clouds, $\mach\sim10-20$ \citep[e.g.,][]{BT07} implying a relatively smaller contribution of turbulent pressure support in the simulation compared to molecular clouds.
Indeed, \citet{KM05} showed that the density threshold for star formation, understood as the density where thermal pressure equals turbulent pressure, in a supersonically turbulent medium is $\rho_{\mathrm{crit}}\sim\rho_0\,\mach^2$.

\subsection{Disc Fragmentation}
\label{sec:disk_fragmentation}

The idea that supersonic turbulence suppresses star formation and fragmentation does not apply in the $\sim1\,\pc$ vicinity of the sink particle where the densities exceed $\sim10^6\,\cc$. In Figure \ref{fig:zoomed_dens_projections} we show that a disc of size $\sim0.1\,\pc$ forms around the sink particle with rotational velocities approaching $\sim15\,\kms$ at $r\sim10^{-2}\,\pc$. It is interesting why this disc does not become unstable and fragment into multiple sink particles, given that disc fragmentation has been seen in recent simulations of Pop III.1 star formation \citep[e.g.,][]{STA10,CLA11a,GRE11}, though at much higher densities and smaller spatial scales than probed here.


\begin{figure}
\includegraphics[scale=0.45, clip, trim = 5 15 20 0   ]{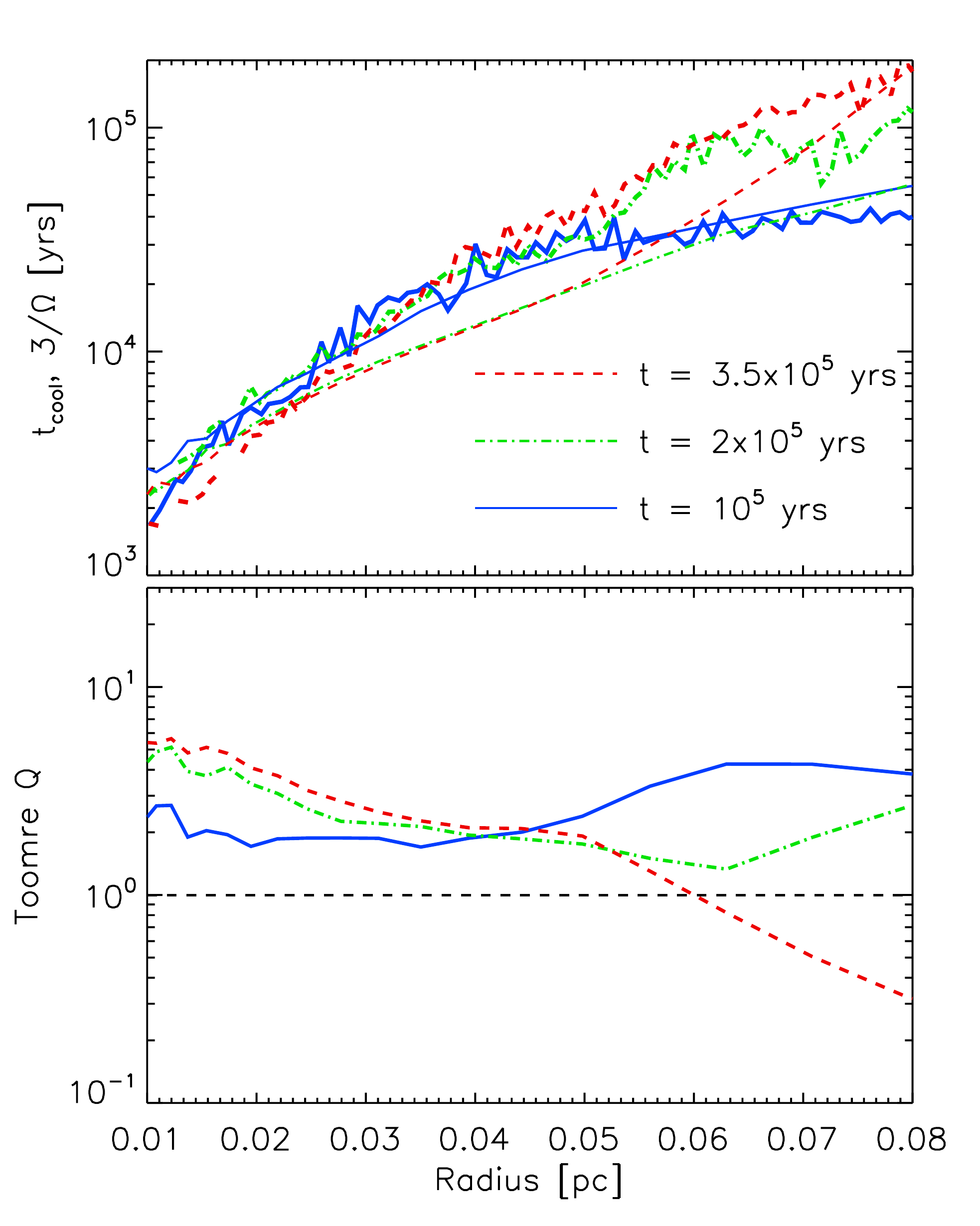}
\caption{Cooling timescale $\tcool$ and $3/\Omega$ (top panel) and the Toomre $Q$ parameter (bottom panel) plotted as a function of distance from the sink particle for gas with $n>10^6\,\cc$ which roughly selects gas in the sink particle's disc. In the top panel, thick lines are the cooling time while the thin lines represent $3/\Omega$. Both panels suggest that the disc is stable against fragmentation. Specifically, in the top panel, $3/\Omega$ is generally below $\tcool$ (the Gammie criterion), while in the bottom panel, $Q\gtrsim1$ throughout the disc's extent. For reference, the disc radius is approximately $\sim0.05-0.07\,\pc$. }
\label{fig:gammie_toomre}
\end{figure}

In general, two conditions must be met for a disc to fragment. The disc must be gravitationally unstable which requires that the Toomre $Q$ parameter, $Q=\cs \Omega / \pi G \Sigma$, be less than unity \citep{TOM64,GL65}, where $\cs$ is the sound speed in the disc, $\Omega=\vrot / R$ is the angular velocity, and $\Sigma$ is the gas surface density. Additionally, even if a disc is gravitationally unstable, a number of effects can prevent fragmentation into gravitationally bound collapsing clumps \citep[e.g.,][]{KM11}. In particular, \citet{GAM01} argued that a key effect influencing the fragmentation behavior of a disc is the balance between cooling and the dissipation of turbulence, which is a source of heating. Quantitatively, Gammie found that if $\tcool<3/\Omega$ then discs tend to fragment into clumps.

In the top panel of Figure \ref{fig:gammie_toomre} we show the mass-weighted average of the cooling time as a function of distance from the sink particle for gas with density $n>10^6\,\cc$ at three different times (thick lines). We also show the mass-weighted average of $3/\Omega$ (thin lines). As is shown, $3/\Omega$ and $\tcool$ trace each other in the inner disc ($r\lesssim0.03\,\pc$), while $3/\Omega$ is a factor of $\sim2$ lower than $\tcool$ at larger radii and particularly in the last two times shown. According to the results of \citet{GAM01}, this disc is stable to fragmentation given its inefficient cooling and relatively rapid rotational velocity. For reference, the radius of the disc is sharp and is approximately $\sim0.05-0.07\,\pc$. 

In the bottom panel of Figure \ref{fig:gammie_toomre}, we show the Toomre $Q$ parameter as a function of distance from the sink particle. To calculate $Q(r)$, we rewrite the gas surface density as $\Sigma(r) = \rho(r)H$, where $H$ is the characteristic vertical thickness of the disc. We have verified that for the disc here, $H\approx0.05\,\pc$ is approximately constant both in radius and time, and that a more detailed calculation of $\Sigma(r)$ would come within a factor of two of this estimate. Thus, $Q(r)$ provides an additional reason for the absence of disc fragmentation since $Q\gtrsim1$ within the disc's extent and actually increases slightly with time.

\section{HD Cooling and Pop III.2 Star Formation}
\label{sec:hd_cooling}

Elevated electron abundances created by the ionizing radiation emitted by the first stars, or in our case, by collisional ionization in $\tvir>10^4\,\kelvin$ haloes, are thought to result in an increase of the HD abundance (see Section \ref{chemistry}). HD is an agent that can act as a coolant in metal-free gas below $200\,\kelvin$. These lower temperatures reduce the Jeans mass and may result in stars with lower characteristic masses, the so-called Pop III.2. However, as discussed in Section \ref{sec:results}, HD is never a significant coolant. Some gas does cool to temperatures $<100\,\kelvin$ as seen in Figure \ref{fig:dens_temp_full}, but this is the result of adiabatic expansion, not HD cooling.
 
\begin{figure}
\includegraphics[scale=0.5, clip, trim =  20 0 0 0  ]{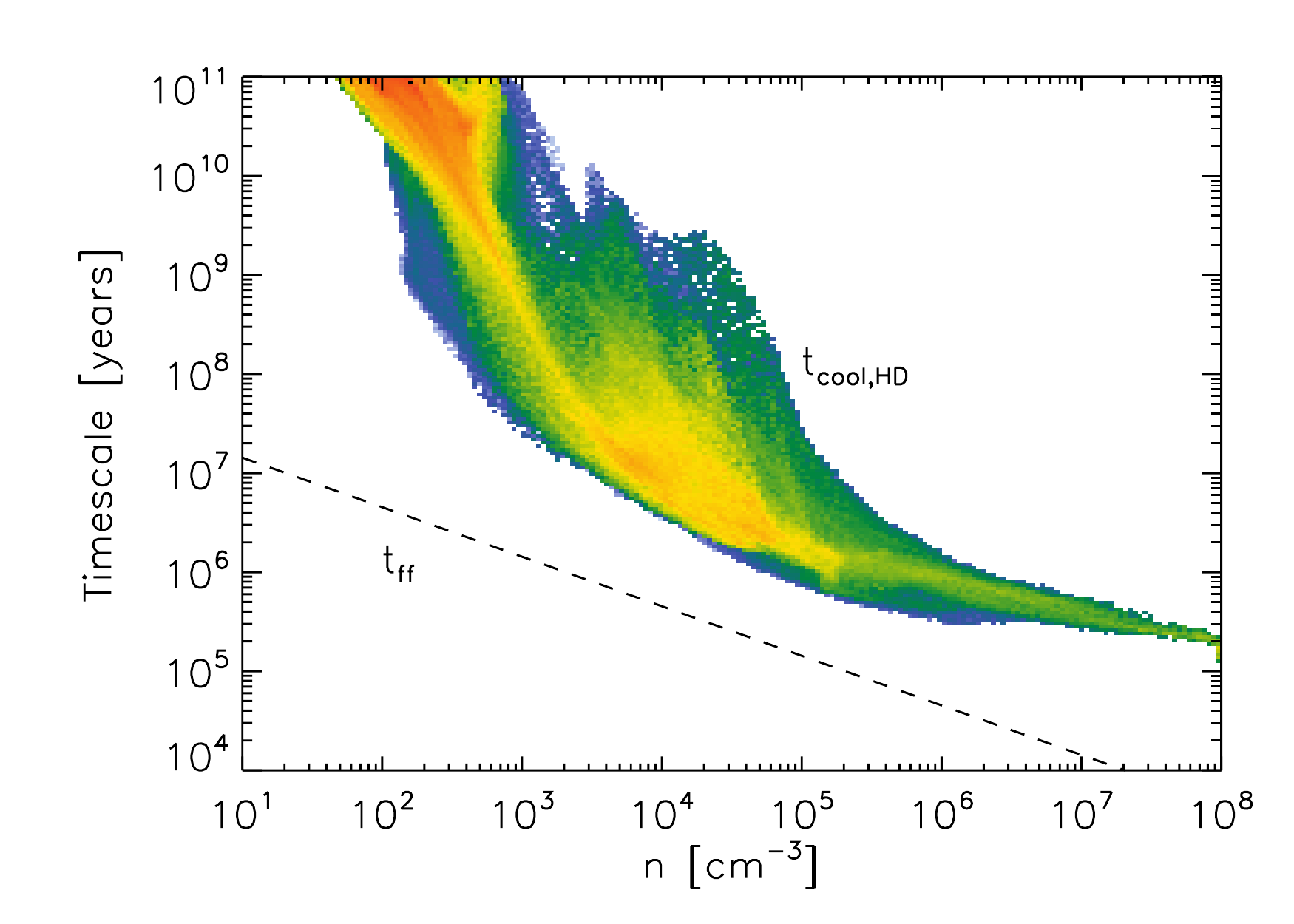}
\caption{HD cooling time and the free-fall time as a function of density when the sink particle forms. For HD cooling to be thermodynamically significant, the HD cooling time must be shorter than the free-fall time, which is never realized. This is a result of many factors, including the reduction in the free electron fraction when $\htwo$ cooling becomes significant at $n\approx200\,\cc$ and the reduction in the HD cooling rate when it thermalizes at $n\approx10^5\,\cc$. }
\label{fig:hd_cooling_time}
\end{figure}

In Figure \ref{fig:hd_cooling_time} we show a phase plot of the HD cooling time, $t_{\mathrm{cool,HD}}$ as a function of density at the time when the sink particle formed. We also show the free-fall time $\tff$. For HD cooling to be significant, it is necessary that $t_{\mathrm{cool,HD}}<\tff$. This criterion is never met.

Many studies have explored the HD cooling mode in Pop III.2 star formation. \citet{JB06} calculated that an HD abundance of $\abundhd\sim10^{-8}$ is needed for HD emission to cool gas to $\tcmb$ in a Hubble time. They suggested that this sets a firm lower limit on the HD abundance needed for its cooling to have a thermodynamic impact. Using one-zone thermodynamic models, they also found that this HD abundance is realized in the downstream of $T\sim10^4\,\kelvin$ virialization shocks, in supernova remnant cooling, and in the collapse of relic Pop III.1 HII regions. In a cosmological setting, \citet{GB08} detected gas with temperatures $\sim\tcmb$, which they attribute to efficient HD cooling, in an atomic cooling halo of similar mass to the one we analyze. \citet{YO07} found that HD cooling was responsible for gas cooling to the CMB temperature after the collapse of a Pop III.1 relic HII region and argued this was responsible for reducing the mass of the clump which eventually will undergo runaway collapse. There has even been a suggestion, by \citet{MCG08}, that in Pop III.1 star formation HD cooling may play a thermodynamically significant role.
In contrast, other studies that have examined the collapse of metal-free gas in a similar context to the one explored here are in agreement that HD never plays a major thermodynamic role \citep{OMU01,OSH08,SHA10}.

We propose two related reasons for the lack of significant HD cooling in our simulation. First, as found by \citet{WG111} using one-zone models, the LW flux necessary to suppress HD cooling is $J_{\mathrm{crit,HD}}\approx 10^{-22}\,\intensunits$. This is far smaller than the flux required to completely suppress $\htwo$ formation and cooling, $J_{\mathrm{crit,H_{2}}}\approx 10^{-20}-10^{-16}\,\intensunits$ \citep{SHA10}. This critical HD flux is not the result of direct HD photodissociation, but is instead due to the partial photodissociation of $\htwo$ which never permits a significant fraction of gas to cool below $\sim300\,\kelvin$ where significant HD fractionation can occur. Second, at $\ncool\approx200\,\cc$ where $\htwo$ cooling becomes effective, the free electron abundance is not high enough to drive up the $\htwo$ abundance. Gas heated to high temperatures in virialization shocks can have a free electron abundance of $\abunde\sim10^{-2}$, but only at low densities, $10^{-2}-10^{-1}\,\cc$. At larger densities around $\ncool$ the electron abundance drops to $\sim10^{-4}$ (see Figure \ref{fig:abundance_multiplot}), below the residual abundance after recombination, simply because the electron recombination time is inversely proportional to density. This would seem to suggest that Pop III.2 star formation is very similar to Pop III.1, at least in the scenario explored here.

%
%

\section{Discussion}
\label{sec:discussion}

\subsection{Lyman-Werner Radiation Field}
\label{sec:lw_radiation_field}

\begin{figure*}
\begin{center}
\includegraphics[scale=0.6, clip, trim = 20 5 5 5 ]{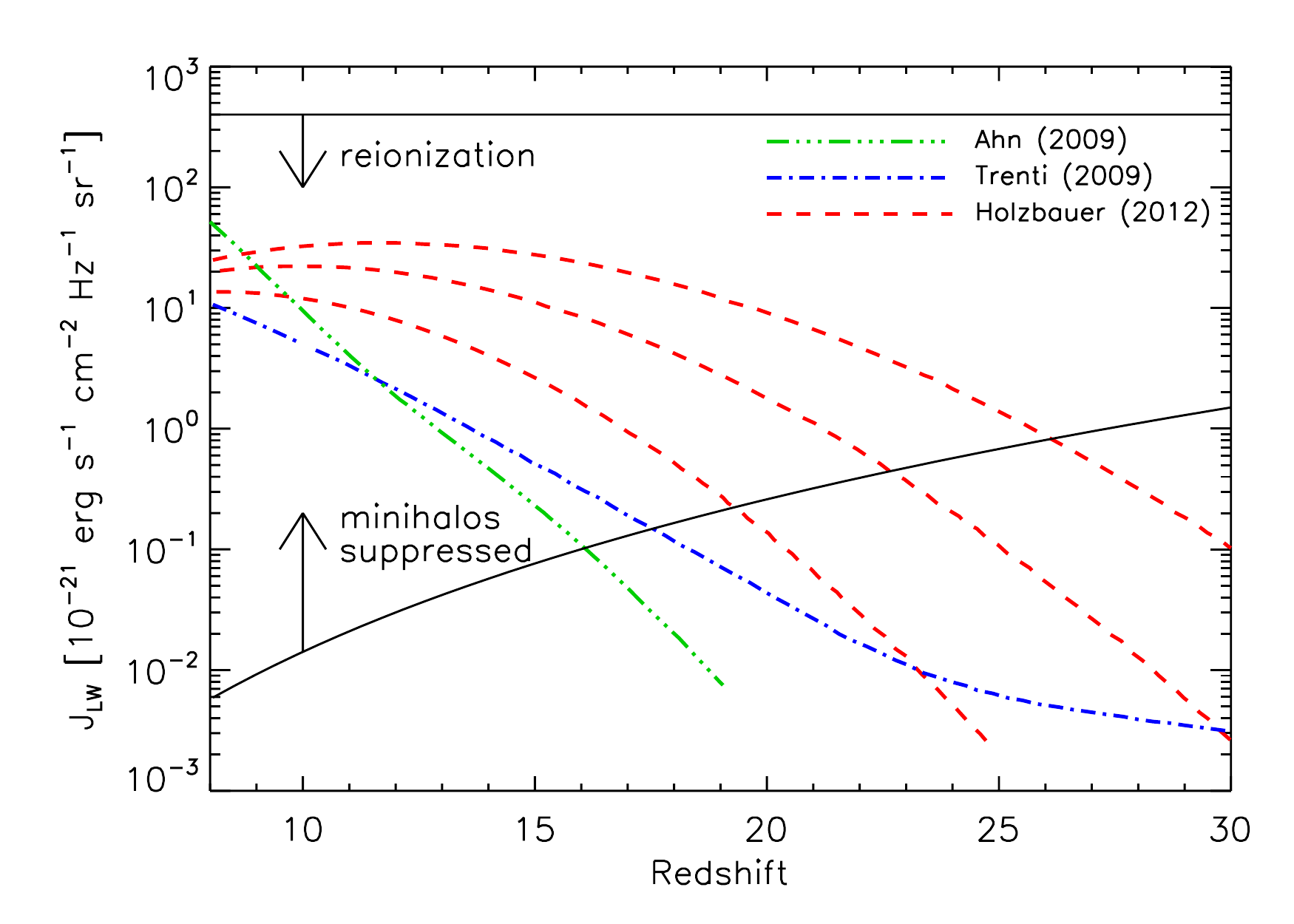}
\caption{Theoretical estimates of the redshift evolution of the average radiation intensity in the LW bands ($\jlw$) compiled from the literature. The curves show the results of \citet{AHN09} (\emph{green triple-dot dashed lined}), \citet{TS09} (\emph{blue dashed-dot line}), and \citet{HF12} (\emph{red dashed lines}). The three curves of \citet{HF12} correspond to different choices for the minimum virial mass of a halo capable of forming stars, which they take, from top to bottom, as $10^6,10^7$, and $10^8\,\solarmass$. The top black line denotes the LW radiation background which accompanies reionization of the IGM at $z=9$, assuming an escape fraction of ionizing radiation of $\fesc=0.1$ and $10$ ionizing photons-per-baryon required for reionization. The bottom black line is an estimate of the LW background needed to suppress cooling in haloes which rely exclusively on $\htwo$. See the text for further details.   }
\label{fig:jlw_compilation}
\end{center}
\end{figure*}

A key physical parameter in this study is the intensity of the constant LW radiation background. Given the large mean free path of photons with energies below $13.6\,\ev$, the formation of the first stars should have established a LW background well before the completion of reionization \citep[e.g.,][]{HAR00}.

In terms of the cosmic star formation rate, we can estimate $\jlw$ as \citep[e.g.,][]{JOH11}
\begin{equation}
\jlw\approx2\left(\frac{\eta_{\mathrm{LW}}}{10^4}\right)\left(\frac{\dot{\rho_{*}}}{10^{-2}\,\solarmass \,\mathrm{yr}^{-1}\,\mathrm{Mpc}^{-3}}\right)\left(\frac{1+z}{10}\right)^3 \mbox{,}
\label{j21_star_formation}
\end{equation}
where $\eta_{\mathrm{LW}}$ is the number of LW photons produced per stellar baryon, $\dot{\rho_{*}}$ is the star formation rate per comoving volume, and the lifetime of stars producing the LW radiation is implicitly assumed to be $5\times10^6\,\yrs$. The star formation rate is normalized with a reasonable value around $z=10$ \citep[e.g.,][]{TS09}.

Another estimate of $\jlw$ can be obtained by considering the ionizing background needed to reionize the Universe. Assuming that reionization was driven by photons with energies just above the Lyman limit, it is straightforward to estimate the value of the UV background just below the limit, $J_{21}^-\approx \jlw$ \citep[e.g.,][]{BL03},
\begin{equation}
J_{21}^-\approx 400\left(\frac{N_{\gamma}}{10}\right)\left(\frac{f_{\mathrm{esc}}}{0.1}\right)^{-1}\left(\frac{1+z}{10}\right)^{3}\mbox{ ,}
\label{eq:j21_reionization}
\end{equation}
where $N_{\gamma}$ is the number of ionizing photons-per-baryon required to reionize the Universe and $f_{\mathrm{esc}}$ is the ionizing photon escape fraction from high-redshift star forming galaxies. Both of the above estimates suggest that before reionization was complete, radiation intensities as high as $\jlw\sim100$, the value we adopt here, were realized in the Universe. Depending on the scenario for reionization, these may overestimate the mean background intensity, but \citet{DIJ08} showed that, due to dark matter halo clustering, a small fraction of haloes can experience LW intensities greatly above the global mean value.

Even though the UV background intensity around the time of reionization is very uncertain, the above estimates suggest it was likely far above the threshold, $\jlw\sim10^{-1}$, that would have suppressed the formation of $\htwo$ in small-mass $\sim10^5-10^6\,\solarmass$ haloes, thus delaying cooling and star formation until the assembly of atomic cooling haloes. As we shall discuss in Section \ref{sec:direct_observations}, the precise value of $\jlw$ may significantly change the results of our simulation, particularly the star formation efficiency.

For more accurate estimates of $\jlw$ we can look to other studies which have undertaken detailed modeling of the redshift evolution of the LW background intensity. In Figure \ref{fig:jlw_compilation}, we show three such estimates in the literature. We also show the LW intensity which would have accompanied reionization at $z=9$ (top solid line) assuming the fiducial values in Equation (\ref{eq:j21_reionization}). Additionally, we plot the redshift-dependent LW radiation intensity $J_{\mathrm{LW, crit}}(z)$ needed to completely suppress $\htwo$ formation in haloes with $\tvir<10^4\,\kelvin$ (bottom solid line). We compute $J_{\mathrm{LW, crit}}(z)$ by equating the $\jlw$-dependent minimum halo mass needed for $\htwo$ cooling with the halo mass at which the virial temperature equals $10^4\,\kelvin$ \citep[equations 12 and 14 from][respectively]{TS09} to obtain
\begin{equation}
J_{\mathrm{LW, crit}}(z) = 1.5\,\left(\frac{1+z}{31}\right)^{4.5}.
\label{eq:jlw_crit}
\end{equation}
We refer the reader to the original references for the details of the models, but briefly summarize the critical components. \citet{TS09} and \citet{HF12} model $\jlw(z)$ in a semi-analytic fashion, using the formalism of \citet{PS74} modified for ellipsoidal collapse by \citet{ST02} to estimate the mass function of collapsed haloes, which they combine with a prescription for star formation in these haloes. \citet{TS09} self-consistently utilize their calculated LW background to estimate the minimum mass of a star-forming halo. The three curves in Figure \ref{fig:jlw_compilation} from \citet{HF12} correspond to different choices for the minimum mass of a star-forming halo; $M_{\mathrm{min}}=10^6$, $10^7$, and $10^8\,\solarmass$ from top to bottom, respectively, each with a fixed star formation efficiency of $0.1$. \citet{AHN09} do not rely on a semi-analytic approach, but instead carry out transfer of ionizing and LW radiation in a large, cosmological, $N$-body simulation. Their resolution, however, is limited to $10^8\,\solarmass$, which may be responsible for their $\jlw$ estimate being the lowest at high redshifts, as can be seen in Figure \ref{fig:jlw_compilation}. 

There are numerous uncertainties in these estimates of $\jlw(z)$, particularly in the choice of star formation efficiency (which is likely to depend on $\jlw$ itself, as well as the redshift, halo mass, and metallicity), LW escape fraction of radiation, mass and multiplicity of Pop III sources, and different prescriptions for star formation feedback. We note that all these models assume that Pop III sources have an extremely top-heavy IMF with a characteristic mass $\sim100\,\solarmass$. In spite of these uncertainties, most of these estimates seem to converge around $z=10$, but they still underpredict the LW intensity that would have accompanied reionization by $z=9$. Additionally, all models suggest that at $z\lesssim15-20$, the average LW radiation field is large enough to completely suppress cooling and star formation in haloes with $\tvir<10^4\,\kelvin$.

Furthermore, as discussed in \citet{AHN09} and \citet{HF12} \citep[see also][]{DIJ08}, the LW radiation field is expected to be very spatially uniform around the time of reionization, in the sense that isolated regions with $\jlw$ significantly above or below the mean were extremely rare. This homogeneity increases with decreasing redshift and increasing spatial scale. Compared to this spatially homogenous LW radiation feedback, chemical enrichment of the IGM by galactic outflows has been suggested to be extremely inhomogeneous \citep[e.g.,][]{SCA02,TOR07,TRE09}. This suggests that the scenario we explore in this work, a chemically pristine atomic cooling halo subjected to a strong LW background, is a physically plausible scenario for the formation of metal-free stellar clusters.

%
%

\subsection{Internal Feedback}
\label{sec:internal_feedback}
This work focused on the impact of an external LW radiation background and its effect on star formation in metal-free atomic cooling haloes. However, once a starburst begins, LW and photoionizing radiation feedback from internal sources will likely impact subsequent star formation. \citet{GB01} explored the conditions under which a metal-free gaseous clump exposed to LW radiation can continue to collapse and form stars. They argue that star formation in  gas clumps will be suppressed if the $\htwo$ photodissociation time, $\tdis$, is shorter than the free-fall time of the clump. \citet{GB01} derived that $\tdis$ for a spherical, homogeneous cloud of density $n$, mass $M$, and distance $D$ from a source which produces $\ndis$ LW photons per second is given by
\begin{equation}
\tdis= 20\abundhtwo n^{2/3}M_{\mathrm{solar}}^{1/3}D_{\mathrm{pc}}^2\fdis^{-1}\fabs^{-1} \left(\frac{\ndis}{10^{48}\,\mathrm{s}^{-1}}\right)^{-1}\,\yrs
\label{eq:t_dis}
\end{equation}
where $\fabs$ is the fraction of incident photons which are actually absorbed by the cloud, $\fdis$ is the fraction of $\htwo$ excitations by a LW photon which result in a successful dissociation, and $M_{\mathrm{solar}}$ and $D_{\mathrm{pc}}$ are measured in units of $\solarmass$ and parsecs, respectively. Considering the simulation here, a clump at a distance of $\sim1\,\pc$ from the source of LW photons (which we take to be the sink particle) has a characteristic mass on order of the Jeans mass $\sim3000\,\solarmass$, density $\sim10^5\,\cc$, and an $\htwo$ abundance $\sim10^{-3}$. If we reasonably assume that the central sink particle forms roughly ten $10\,\solarmass$ stars (see Section \ref{sec:fragmentation}) the LW photon production rate is $\ndis\approx10^{48}\,\mathrm{s}^{-1}$. The resulting $\htwo$ photodissociation time $t_{\mathrm{dis}} \approx 6\times10^4\,\yrs $ assuming that $\fdis=\fabs=0.1$. The free-fall time is $\tff(10^5\,\cc)\approx1.5\times10^5\,\yrs$, therefore this clump will have its primary cooling agent, $\htwo$, photodissociated and thus its further collapse suppressed. 


 \begin{figure}
 \begin{center}
\includegraphics[scale=0.52, clip, trim = 30 5 5 5 ]{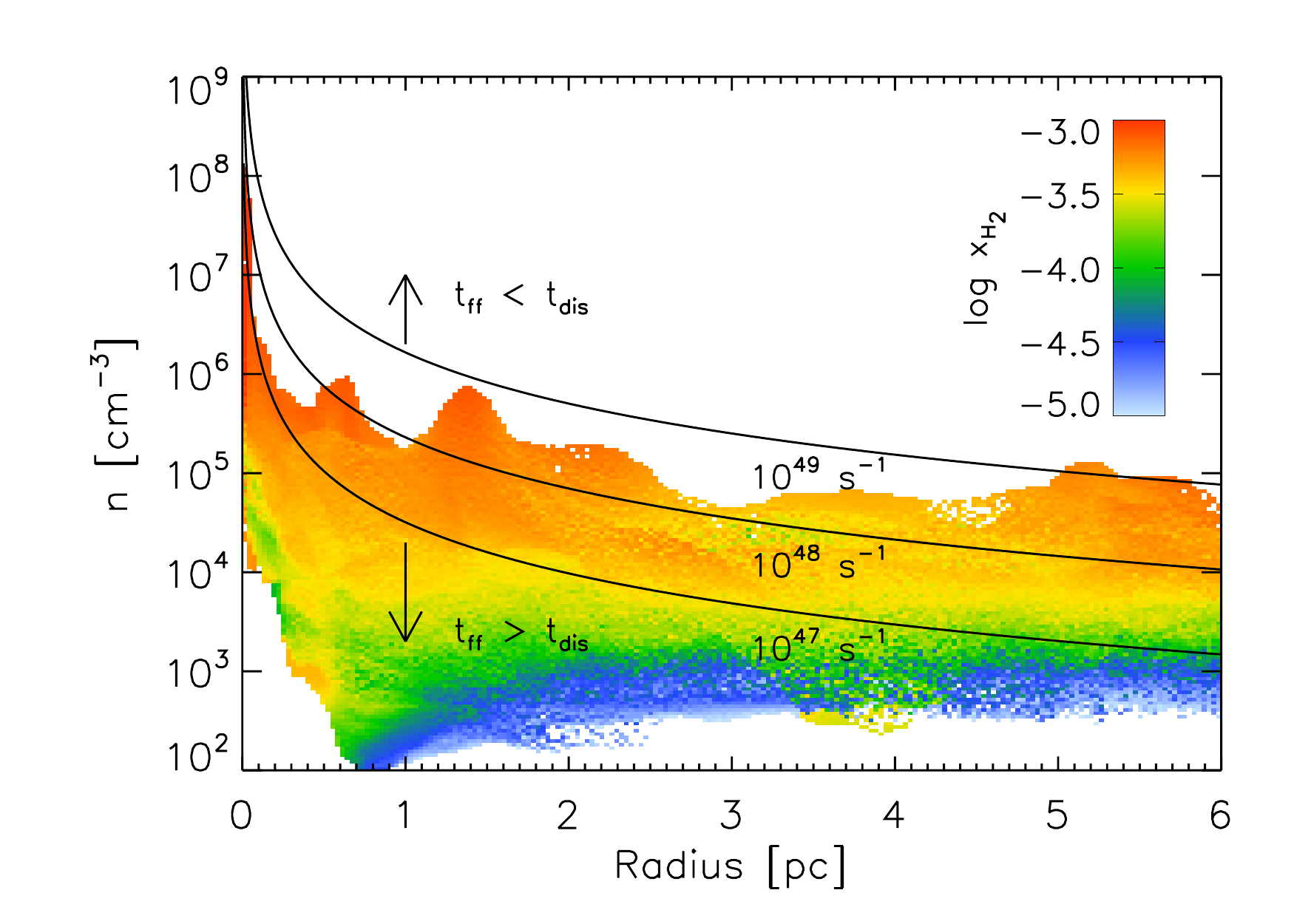}
\caption{Gas density as a function of distance from the sink particle $10^5$ years after sink particle formation colour coded by the mass-weighted $\htwo$ abundance. The required density for a given clump's free-fall time to equal its $\htwo$ dissociation time is also shown by the solid lines, that were computed assuming all clumps have a mass of $3000\,\solarmass\approx\mj$ and and $\htwo$ abundance of $\abundhtwo=10^{-3}$. The different solid lines correspond to three different choices for the LW photon production rate $\ndis$: $10^{49}\,\mathrm{s}^{-1}$ (top) is appropriate for a single $100\,\solarmass$ star, $10^{48}\,\mathrm{s}^{-1}$ (middle) for ten $10\,\solarmass$ stars, while $10^{47}\,\mathrm{s}^{-1}$ (bottom) would be the rate from ten $5\,\solarmass$ stars. Assuming a homogeneous clump, its density must lie above a solid line for its collapse to occur before its supply of $\htwo$ is exhausted by photodissociation.  }
\label{fig:radial_density_lw}
\end{center}
\end{figure}

In reality, the effect of LW feedback on nearby collapsing clumps will depend strongly on their distances from the radiation source and on the level of pre-condensation in the gas, i.e., how dense the clump is when the neighboring LW source turns on. In Figure \ref{fig:radial_density_lw} we show the radius and density dependence of the $\htwo$ abundance in cells with $\abundhtwo>10^{-5}$ at a time $10^5$ years after sink particle formation, roughly the Kelvin-Helmholtz time of massive stars that would be the primary producers of LW photons. The cells are colour coded by their $\htwo$ abundance. We also show the density where $\tff=\tdis$ for three different choices of the LW photon production rate. To derive this density, we assume that collapsing clumps have a total mass of $3000\,\solarmass$,\footnote[1]{We note that the clumps visible in Figure \ref{fig:radial_density_lw} at $\sim0.5\,\pc$ and $\sim1.5\,\pc$ are in fact not Jeans unstable and have masses well below $3000\,\solarmass$. This downward mass revision will decrease their $\htwo$ photodissociation time which depends weakly on mass, $\tdis\propto M^{1/3}$. } $\htwo$ abundance of $10^{-3}$, and that $\fdis=\fabs=0.1$. Evidently the effect of LW radiative feedback on subsequent star formation is a strong function of $\ndis$ and the source distance. If star formation which occurs in the sink particle results in a LW photon production rate of $\ndis\approx10^{48}\,\mathrm{s}^{-1}$, it is clear from Figure \ref{fig:radial_density_lw} that further collapse and star formation will be strongly suppressed in the uncollapsed fraction of the self-shielding core, even given the large simplifications we have made in this analysis. This may have an adverse effect on the overall star formation efficiency and detectability of these systems.

%
%

\subsection{Direct JWST Observations of a Metal-Free Stellar Population}
\label{sec:direct_observations}

Since the basic characteristics and formation mechanisms of the first cosmic structures are very uncertain, one of the main goals of future observations with the JWST is to observe light from the first stars and galaxies. By modeling metal-free synthetic stellar spectra, \citet{ZAK11} determined JWST exposure limits and colour criteria for high-redshift, metal-free galaxies. Given the uncertainties of these sources, \citet{ZAK11} explored a large parameter space defined by the shape of the stellar IMF, nebular emission strength, stellar population age, stellar mass, and formation redshift. Their most optimistic scenario involves a $3$ Myr old stellar population with an extremely top-heavy IMF ($dn/dM \propto M^{-2.35}$ for $50<M/\solarmass<500$) and maximal nebular emission, requiring a vanishing escape fraction of ionizing radiation, $\fesc=0$. At $z=10$, a cluster with stellar mass $M_{*}\approx10^5\,\solarmass$ and the above characteristics would just be detectable (at $10\sigma$) with JWST in ultra-deep ($100$ hr) broadband exposures.

Given our finding that $\approx1.5\times10^4\,\solarmass$ of cold gas becomes available for star formation in an atomic cooling halo at $z\approx12.1$ exposed to a LW intensity of $J_{21}=100$, it seems extremely unlikely this starburst could be detected by JWST, even if $100\%$ of the mass turns into stars with an extremely top-heavy IMF. It is conceivable, however, that a weaker UV background could still have suppressed star formation until the assembly of an atomic cooling halo, but could have allowed for a larger global star formation efficiency. An inspection of Figure \ref{fig:jlw_compilation} suggests that a background LW intensity $\jlw\gtrsim1$, at least for $z\lesssim30$, would have fully suppressed star formation in halos not capable of atomic cooling. 

To get a sense of how the mass of cold gas depends on $J_{21}$, we proceed as follows.
First, recall Equation (\ref{eq:ncool2}) which expresses the density at which $\htwo$ cooling becomes effective, $\ncool$, as a function of $J_{21}$ and $\abunde$. This cooling density partially determines the mass of cold $T\lesssim10^3\,\kelvin$ gas, $\mcold$, available for the first burst of star formation. Once a halo reaches the atomic cooling limit, the accretion rate of gas onto the self-shielding core, $\dot{M}_{\mathrm{cold}}$, will not a strong function of $J_{21}$. With this, we can estimate the cold gas mass as $\mcold \approx \dot{M}_{\mathrm{cold}} \, \tff(n_{\mathrm{cool}}) \propto J_{21}^{-1/3}\,\abunde^{1/3}$. The abundance of free electrons would be higher at lower densities, but by how much is difficult to determine given the highly non-equilibrium electron abundance. For example, if $\abunde$ is $10$ times higher than in our simulation at $\ncool$ and if $J_{21}=10$, the mass of cold gas would be $\mcold\sim10^{2/3}\times1.5\times10^4\,\solarmass\approx7\times10^{4}\,\solarmass$. If we furthermore assume a relatively high star formation efficiency, $f_{*}=0.1$, which represents the fraction of cold gas that will eventually turn into stars, we find a total stellar mass of $M_{*}\sim f_{*}\,\mcold\sim7000\,\solarmass$. Even if these stars form with an extremely top-heavy IMF, a possibility recent simulations suggest is unlikely \citep[e.g.,][]{STA10,CLA11a,GRE11}, the analysis of \citet{ZAK11} implies this cluster will not be detectable with the JWST.

One effect, however, which may render these targets within the JWST detection limit is gravitational lensing. Low-redshift galaxy clusters are capable of magnifying high-redshift sources by factors of $\mu\sim10-100$. Indeed, the recently discovered $z\approx9.6$ galaxy MACS 1149-JD1 \citep{ZH12} is gravitationally lensed by a foreground galaxy cluster with a magnification factor of $\mu\approx14$, though lensing was not strictly required for its detection. To this end, \citet{ZAK12} estimated the conditions under which the JWST could detect gravitationally lensed, metal-free stellar populations behind the $z\approx0.546$ galaxy cluster MACS J0717.5+3745, an ideal gravitational lens for high-redshift stellar sources. Mock halo catalogs of metal-free galaxies, generated with the methodology of \citet{TRE09}, were simulated to lie beyond this particular galaxy cluster and the expected number of metal-free stellar populations as a function of limiting magnitude was generated.
The primary source of uncertainty in these estimates is $\epsilon$, the efficiency of Pop III.2 star formation, defined as $M_{\mathrm{*,popIII.2}} = \epsilon M_{\mathrm{halo}}(\omegab/\omegam)$. This efficiency could be broken down as $\epsilon = f_{*}f_{\mathrm{cold}}$, where $f_{\mathrm{cold}}$ is the fraction of a halo's baryonic mass which is cold and available for star formation and $f_{*}$ is, as before, the fraction of cold gas that will eventually become incorporated into stars. 

\citet{ZAK12} find that with efficiencies as low as $\epsilon\approx10^{-3}$, clusters of metal-free stars with moderately top-heavy IMFs, $M_{\mathrm{char}}=10\,\solarmass$, will be visible in deep JWST lensing surveys around MACS J0717.5+3745. Our optimistic $J_{21}=10$ scenario, described above, results in an efficiency $\epsilon\approx(7000\,\solarmass/3\times10^7\solarmass)(\omegam/\omegab)\approx10^{-2.9}$. Thus, according to the estimates of \citet{ZAK12} and our relatively crude estimate for the Pop III.2 star formation efficiency, it appears that JWST will marginally detect metal-free stellar populations in deep, foreground cluster lensing enhanced surveys. This conclusion, though, is very approximate and more detailed studies are needed to fully assess the observational prospects for detecting these primitive, low luminosity stellar clusters.

%
%

\section{Summary and Conclusions}
\label{sec:summary}

We have performed a high-resolution cosmological simulation focused on the collapse of metal-free gas in a region of the Universe subjected to a strong molecule-dissociating Lyman-Werner background. We resolved densities up to $10^8\,\cc$ and length scales down to $\sim1000$ AU. This simulation utilized a fully non-equilibrium primordial chemistry network coupled with a non-local column density calculation to accurately compute the $\htwo$ and HD photodissociation rates. Additionally, our use of sink particles allowed us to evolve the simulation for many free-fall times past the first gravitational collapse.

Here we summarize the main findings of this work:

\begin{itemize}

\item With a LW intensity $J_{21} = 100$ in a $1\,\Mpc ^3$ (comoving) box, effective $\htwo$ cooling first occurs in a $3\times10^7\,\solarmass$ halo at $z\approx12.1$. This cooling results in a thermal instability and a cold, dense core develops in which $\htwo$ shields itself from LW radiation. Upon sink particle creation at $n=10^8\,\cc$, the central core containing self-shielding, cold gas has a mass of $\sim10^4\,\solarmass$, characteristic size $\sim5\,\pc$, temperature $\sim400\,\kelvin$, mean rms velocity $\vrms\approx6\,\kms$, and Mach number $\approx2$. Mach numbers increase to $\approx2-4$ within $3.5\times10^5$ years.

\item Within the inner $\sim10\,\pc$ of the halo, the gas flow becomes supersonic only when $\htwo$ cooling becomes effective. The gas density PDF in the cold, self-shielding gas is approximately log-normal and has a width consistent with the turbulent Mach number of the flow given expectations derived from idealized simulations in which turbulence is driven by a mixture of solenoidal and compressive forcing. With time, the density PDF acquires a high density power-law tail as self-gravitating gas decouples from the turbulent flow. 

\item The rate of gaseous collapse on scales comparable to the size of the self-shielding region is suppressed compared to what would be expected if all the gas in the self-shielding core was collapsing at its free-fall rate. By the time the sink particle grows to $\sim6\%$ of the total mass of the self-shielding cloud, no other sinks have formed, suggesting any additional sites of fragmentation remain unresolved in the simulation. We find an upper-limit to the star formation rate per free-fall time of $\sfrff\approx0.1$ on scales of $\sim10\,\pc$. We argue this is due to the additional effective pressure provided by infall-driven supersonic turbulence on large scales and centrifugal support on smaller ones. 

\item HD cooling was found to be thermodynamically insignificant for the entirety of the gas evolution. We attribute this to partial photodissociation of $\htwo$ and the low free electron fraction when $\htwo$ cooling becomes significant. Given the relatively low LW intensity needed for HD formation to be suppressed, $\jlw\sim0.1$, in comparison with the relatively high intensity required to delay gaseous collapse until the formation of atomic cooling haloes, $\jlw\sim10$, it seems unlikely that HD cooling is ever significant in Pop III.2 star formation, at least in the scenario explored here.

\item Given the similarities between the thermodynamic behavior of Pop III forming halos with and without LW feedback, and the complete absence of significant HD cooling in both cases, we suggest that Pop III.2 star formation delayed by LW radiation is very similar to Pop III.1 star formation. Additionally, the similarities of our results with other simulations that probed higher densities and smaller spatial scales \citep[e.g.,][]{STA10,CLA11a,GRE11} seem to suggest that Pop III.2 stellar masses in the LW-delayed mode are likely comparable to Pop III.1 masses, $\sim1-40\,\solarmass$, possibly even lower.

\end{itemize}

\section*{Acknowledgments}

CSS is grateful to Paul Ricker, Eiichiro Komatsu, John Wise, Erik Zackrisson, Jeremy Ritter, and many others for enlightening conversations. CSS also thanks Robi Banerjee and Chris Lindner for providing software used to produce some of the visualizations here. The FLASH code was in part developed by the DOE- supported Alliance Center for Astrophysical Thermonuclear Flashes (ACS) at the University of Chicago. The authors acknowledge the Texas Advanced Computing Center (TACC) at The University of Texas at Austin for providing HPC resources that have contributed to the research results reported within this paper. This study was supported in part by NSF grants AST-0708795 and AST-1009928, by the NASA grants NNX08AL43G and NNX09AJ33G, and by support provided by the Texas Cosmology Center (TCC). CF acknowledges funding provided by the Australian Research Council under the Discovery Projects Scheme (grant no. DP110102191). This research has made use of NASA's Astrophysics Data System.

\bsp

\label{lastpage}


\begin{thebibliography}{}

\bibitem[Abel et al.(1997)]{AB97} Abel, T., Anninos, P., 
Zhang, Y., \& Norman, M.~L.\ 1997, NewA, 2, 181 

\bibitem[Abel et al.(2000)]{ABN00} Abel, T., Bryan, G.~L.,
\& Norman, M.~L.\ 2000, \apj, 540, 39

\bibitem[Ahn \& Shapiro(2007)]{AS07}Ahn, K., \& Shapiro, P.R. 2007,
MNRAS, 375, 881

\bibitem[Ahn et al.(2009)]{AHN09} Ahn, K., Shapiro, P.~R., 
Iliev, I.~T., Mellema, G., \& Pen, U.-L.\ 2009, \apj, 695, 143

\bibitem[Anninos 
\& Norman(1996)]{AN96} Anninos, P., \& Norman, M.~L.\ 1996, \apj, 460, 556

\bibitem[Bader \& Deuflhard(1983)]{BD83}
Bader, G., \& Deuflhard, P.\ 1983, Numer.\ Math.\ 41, 373

\bibitem[Barkana
\& Loeb(2001)]{BL01} Barkana, R., \& Loeb, A.\ 2001, Phys. Rep., 349, 125

\bibitem[Bate et al.(1995)]{BBP95} Bate, M.~R., Bonnell, 
I.~A., \& Price, N.~M.\ 1995, \mnras, 277, 362

\bibitem[Bate et al.(2003)]{BBB03} Bate, M.~R., Bonnell, 
I.~A., \& Bromm, V.\ 2003, \mnras, 339, 577 

\bibitem[Beers 
\& Christlieb(2005)]{BC05} Beers, T.~C., \& Christlieb, N.\ 2005, \araa, 43, 5

\bibitem[Begelman et al.(2006)]{BEG06} Begelman, M.~C., 
Volonteri, M., \& Rees, M.~J.\ 2006, \mnras, 370, 289 

\bibitem[Bergin 
\& Tafalla(2007)]{BT07} Bergin, E.~A., \& Tafalla, M.\ 2007, \araa, 45, 339

\bibitem[Bertschinger(2001)]{BERT01} Bertschinger, E.\ 2001, 
\apjs, 137, 1 

\bibitem[Bonazzola et 
al.(1987)]{BON87} Bonazzola, S., Heyvaerts, J., Falgarone, E., Perault, M., \& Puget, J.~L.\ 1987, \aap, 172, 293 

\bibitem[Bonnor(1956)]{BO56} Bonnor, W.~B.\ 1956, \mnras, 
116, 351 

\bibitem[Bromm et al.(2001)]{BK01} Bromm, V., Kudritzki, 
R.~P., \& Loeb, A.\ 2001a, \apj, 552, 464 

\bibitem[Bromm et al.(2002)]{BCL02}Bromm, V., Coppi, P.S., \& Larson,
R.B. 2002, ApJ, 564, 23

\bibitem[Bromm 
\& Loeb(2003)]{BL03} Bromm, V., \& Loeb, A.\ 2003b, \apj, 596, 34 

\bibitem[Bromm 
\& Larson(2004)]{BL04} Bromm, V., \& Larson, R.~B.\ 2004, \araa, 42, 79 

\bibitem[Bromm et al.(2009)]{BYHM09} Bromm, V., Yoshida, N.,
Hernquist, L., \& McKee, C.~F.\ 2009, \nat, 459, 49

\bibitem[Bromm 
\& Yoshida(2011)]{BY11} Bromm, V., \& Yoshida, N.\ 2011, \araa, 49, 373 

\bibitem[Cen(1992)]{CEN92} Cen, R.\ 1992, \apjs, 78, 341

\bibitem[Chandrasekhar(1951)]{CHAN1951} Chandrasekhar, S.\ 1951, 
Royal Society of London Proceedings Series A, 210, 26 

\bibitem[Cho 
\& Kim(2011)]{CK11} Cho, W., \& Kim, J.\ 2011, \mnras, 410, L8  

\bibitem[Christlieb et al.(2002)]{CHR02}Christlieb, N., et
al. 2002, Nature, 419, 904

\bibitem[Ciardi et al.(2000)]{CIA00} Ciardi, B., Ferrara, A., 
\& Abel, T.\ 2000, \apj, 533, 594 

\bibitem[Ciardi 
\& Ferrara(2005)]{CF05} Ciardi, B., \& Ferrara, A.\ 2005, \ssr, 116, 625 

\bibitem[Clark et al.(2011b)]{CLA11b} Clark, P.~C., Glover, 
S.~C.~O., Smith, R.~J., et al.\ 2011, Science, 331, 1040

\bibitem[Clark et al.(2011a)]{CLA11a} Clark, P.~C., Glover, 
S.~C.~O., Klessen, R.~S., \& Bromm, V.\ 2011, \apj, 727, 110 

\bibitem[Colella 
\& Woodward(1984)]{CW84} Colella, P., \& Woodward, P.~R.\ 1984, Journal of Computational Physics, 54, 174 

\bibitem[Collins et al.(2011)]{COL11} Collins, D.~C., Padoan, 
P., Norman, M.~L., \& Xu, H.\ 2011, \apj, 731, 59 

\bibitem[Couchman 
\& Rees(1986)]{CR86} Couchman, H.~M.~P., \& Rees, M.~J.\ 1986, \mnras, 221, 53

\bibitem[Dekel
\& Birnboim(2006)]{DB06} Dekel, A., \& Birnboim, Y.\ 2006, \mnras, 368, 2

\bibitem[Dekel et al.(2009)]{DEK09} Dekel, A., et al.\ 2009,
\nat, 457, 451

\bibitem[Dijkstra et al.(2008)]{DIJ08}Dijkstra, M., Haiman, Z., Mesinger, A., \&
Wyithe, J.~S.~B. 2008, \mnras, 391, 1961

\bibitem[Draine 
\& Bertoldi(1996)]{DB96} Draine, B.~T., \& Bertoldi, F.\ 1996, \apj, 468, 269 

\bibitem[Dubey et al.(2008)]{DUB08} Dubey, A., Fisher, R., 
Graziani, C., et al.\ 2008, Numerical Modeling of Space Plasma Flows, 385, 
145 

\bibitem[Dubey et al.(2009)]{DUB09} Dubey, A., Reid, L.~B., 
Weide, K., et al.\ 2009, Parallel Computing, 35, 512

\bibitem[Dubey et al.(2011)]{DUB11} Dubey, A., Antypas, K., \& Daley, C.\ 2011, Parallel Computing, 37, 2, ISSN 0167-8191

\bibitem[Dubey et al.(2012)]{DUB12} Dubey, A., Daley, C., 
ZuHone, J., et al.\ 2012, \apjs, 201, 27 

\bibitem[Dunlop(2012)]{DUN12} Dunlop, J.~S.\ 2012, 
arXiv:1205.1543 

\bibitem[Ebert(1955)]{EB55} Ebert, R.\ 1955, \zap, 37, 217 


\bibitem[Eisenstein 
\& Hut(1998)]{EIS98} Eisenstein, D.~J., \& Hut, P.\ 1998, \apj, 498, 13

\bibitem[Elmegreen 
\& Scalo(2004)]{ES04} Elmegreen, B.~G., \& Scalo, J.\ 2004, \araa, 42, 211

\bibitem[Evans et al.(2009)]{EVA09} Evans, N.~J., II, Dunham, 
M.~M., J{\o}rgensen, J.~K., et al.\ 2009, \apjs, 181, 321

\bibitem[Federrath et al.(2008)]{FED08a} Federrath, C., 
Glover, S.~C.~O., Klessen, R.~S., 
\& Schmidt, W.\ 2008, Physica Scripta Volume T, 132, 014025 

\bibitem[Federrath et al.(2008)]{FED08} Federrath, C., 
Klessen, R.~S., \& Schmidt, W.\ 2008, \apjl, 688, L79

\bibitem[Federrath et al.(2010)]{FED10} Federrath, C., 
Banerjee, R., Clark, P.~C., \& Klessen, R.~S.\ 2010, \apj, 713, 269 

\bibitem[Federrath et al.(2010)]{FED10a} Federrath, C., Roman-Duval, J., Klessen, R.~S., Schmidt, W., \& Mac Low, M.-M.\ 2010, \aap, 512, A81

\bibitem[Federrath et al.(2011)]{FED11} Federrath, C., Sur, 
S., Schleicher, D.~R.~G., Banerjee, R., 
\& Klessen, R.~S.\ 2011, \apj, 731, 62 

\bibitem[Ferrara(1998)]{FER98} Ferrara, A.\ 1998, \apjl, 499, 
L17

\bibitem[Flower et al.(2000)]{FL02} Flower, D.~R., Le 
Bourlot, J., Pineau des For{\^e}ts, G., 
\& Roueff, E.\ 2000, \mnras, 314, 753 

\bibitem[Frebel et al.(2005)]{FRE05}Frebel, A., et al. 2005,
Nature, 434, 871

\bibitem[Frebel 
\& Bromm(2010)]{FB10} Frebel, A., \& Bromm, V.\ 2010, arXiv:1010.1261 

\bibitem[Frisch(1995)]{FRI95} Frisch, U.\ 1995, 
Turbulence.~The Legacy of A.~N.~Kolmogorov. By Uriel Frisch. Cambridge 
University Press, 1995. ISBN: 0-521-45103-5

\bibitem[Fryxell et al.(2000)]{FRY00} Fryxell, B., Olson, K., 
Ricker, P., et al.\ 2000, \apjs, 131, 273 

\bibitem[Fumagalli et al.(2011)]{FU11} Fumagalli, M., 
O'Meara, J.~M., \& Prochaska, J.~X.\ 2011, arXiv:1111.2334

\bibitem[Furlanetto 
\& Loeb(2005)]{FUR05} Furlanetto, S.~R., \& Loeb, A.\ 2005, \apj, 634, 1 

\bibitem[Gammie(2001)]{GAM01} Gammie, C.~F.\ 2001, \apj, 553, 
174 

\bibitem[Galli 
\& Palla(1998)]{GP98} Galli, D., \& Palla, F.\ 1998, \aap, 335, 403

\bibitem[Gardner et al.(2006)]{GAR06} Gardner, J.~P., Mather, 
J.~C., Clampin, M., et al.\ 2006, \ssr, 123, 485 

\bibitem[Girichidis et al.(2011)]{GIR11} Girichidis, P., 
Federrath, C., Banerjee, R., \& Klessen, R.~S.\ 2011, \mnras, 413, 2741 

\bibitem[Glover 
\& Brand(2001)]{GB01} Glover, S.~C.~O., \& Brand, P.~W.~J.~L.\ 2001, \mnras, 321, 385

\bibitem[Glover 
\& Mac Low(2007)]{GM07} Glover, S.~C.~O., \& Mac Low, M.-M.\ 2007, \apjs, 169, 239 

\bibitem[Glover et al.(2010)]{GLO10} Glover, S.~C.~O., 
Federrath, C., Mac Low, M.-M., \& Klessen, R.~S.\ 2010, \mnras, 404, 2 

\bibitem[Glover(2011)]{G11} Glover, S.~C.~O.\ 2011, IAU 
Symposium, 280, 313 

\bibitem[Goldreich 
\& Lynden-Bell(1965)]{GL65} Goldreich, P., \& Lynden-Bell, D.\ 1965, \mnras, 130, 97 

\bibitem[Gnedin(2010)]{GNE10} Gnedin, N.~Y.\ 2010, \apjl, 
721, L79 

\bibitem[Greif \& Bromm(2006)]{GB06} Greif, T.~H., \& Bromm, V.\ 2006,
\mnras, 373, 128

\bibitem[Greif et al.(2008)]{GB08} Greif, T.~H., Johnson,
J.~L., Klessen, R.~S., \& Bromm, V.\ 2008, \mnras, 387, 1021

\bibitem[Greif et al.(2009)]{GRE09} Greif, T.~H., Johnson, 
J.~L., Klessen, R.~S., \& Bromm, V.\ 2009, \mnras, 399, 639 

\bibitem[Greif et al.(2010)]{GRE10} Greif, T.~H., Glover, 
S.~C.~O., Bromm, V., \& Klessen, R.~S.\ 2010, \apj, 716, 510 

\bibitem[Greif et al.(2011)]{GRE11} Greif, T.~H., Springel, 
V., White, S.~D.~M., et al.\ 2011, \apj, 737, 75

\bibitem[Greif et al.(2012)]{GRE12} Greif, T.~H., Bromm, V., 
Clark, P.~C., et al.\ 2012, \mnras, 3229 

\bibitem[Haiman et al.(1996)]{HAI96} Haiman, Z., Thoul, 
A.~A., \& Loeb, A.\ 1996, \apj, 464, 523 

\bibitem[Haiman et al.(1997)]{HAI97} Haiman, Z., Rees, M.~J., 
\& Loeb, A.\ 1997, \apj, 476, 458 


\bibitem[Haiman et al.(2000)]{HAR00} Haiman, Z., Abel, T., 
\& Rees, M.~J.\ 2000, \apj, 534, 11 

\bibitem[Heger et al.(2003)]{HEG03} Heger, A., Fryer, C.~L., 
Woosley, S.~E., Langer, N., \& Hartmann, D.~H.\ 2003, \apj, 591, 288 

\bibitem[Hennebelle 
\& Chabrier(2008)]{HC08} Hennebelle, P., \& Chabrier, G.\ 2008, \apj, 684, 395 

\bibitem[Hennebelle 
\& Chabrier(2011)]{HC11} Hennebelle, P., \& Chabrier, G.\ 2011, \apjl, 743, L29 

\bibitem[Holzbauer 
\& Furlanetto(2012)]{HF12} Holzbauer, L.~N., \& Furlanetto, S.~R.\ 2012, \mnras, 419, 718 

\bibitem[Hosokawa et al.(2011)]{HO11} Hosokawa, T., Omukai, 
K., Yoshida, N., \& Yorke, H.~W.\ 2011, Science, 334, 1250 

\bibitem[Hummel et al.(2011)]{HUM11} Hummel, J., Pawlik, A., 
Milosavljevic, M., \& Bromm, V.\ 2012, \apj, in press (arXiv:1112.5207)

\bibitem[Inoue(2011)]{INO11} Inoue, A.~K.\ 2011, \mnras, 415, 
2920 

\bibitem[Johnson \& Bromm(2006)]{JB06}Johnson, J. L., \& Bromm, V.
2006, MNRAS, 366, 247

\bibitem[Johnson et al.(2008)]{JOH08} Johnson, J.~L., Greif,
T.~H., \& Bromm, V.\ 2008, \mnras, 388, 26

\bibitem[Johnson(2011)]{JOH11} Johnson, J.~L.\ 2011, 
arXiv:1105.5701 

\bibitem[Kainulainen et 
al.(2009)]{KAI09} Kainulainen, J., Beuther, H., Henning, T., \& Plume, R.\ 2009, \aap, 508, L35 

\bibitem[Karlsson et al.(2011)]{KAR11} Karlsson, T., Bromm, 
V., \& Bland-Hawthorn, J.\ 2011, arXiv:1101.4024 

\bibitem[Kennicutt 
\& Evans(2012)]{KV12} Kennicutt, R.~C., Jr., \& Evans, N.~J., II 2012, arXiv:1204.3552

\bibitem[Klessen(2000)]{KLE00} Klessen, R.~S.\ 2000, \apj, 
535, 869

\bibitem[Klessen 
\& Hennebelle(2010)]{KH10} Klessen, R.~S., \& Hennebelle, P.\ 2010, \aap, 520, A17 

\bibitem[Kolmogorov(1941)]{KOL41} Kolmogorov, A.\ 1941, 
Akademiia Nauk SSSR Doklady, 30, 301

\bibitem[Komatsu et al.(2011)]{KOM11} Komatsu, E., Smith, 
K.~M., Dunkley, J., et al.\ 2011, \apjs, 192, 18 

\bibitem[Kratter 
\& Murray-Clay(2011)]{KM11} Kratter, K.~M., \& Murray-Clay, R.~A.\ 2011, \apj, 740, 1 

\bibitem[Kritsuk 
\& Norman(2002)]{KN02} Kritsuk, A.~G., \& Norman, M.~L.\ 2002, \apjl, 569, L127 

\bibitem[Kritsuk et al.(2007)]{KRI07} Kritsuk, A.~G., Norman, 
M.~L., Padoan, P., \& Wagner, R.\ 2007, \apj, 665, 416 

\bibitem[Kritsuk et al.(2011)]{KRI11} Kritsuk, A.~G., Norman, 
M.~L., \& Wagner, R.\ 2011, \apjl, 727, L20 

\bibitem[Krumholz et al.(2004)]{KRU04} Krumholz, M.~R., 
McKee, C.~F., \& Klein, R.~I.\ 2004, \apj, 611, 399

\bibitem[Krumholz 
\& McKee(2005)]{KM05} Krumholz, M.~R., \& McKee, C.~F.\ 2005, \apj, 630, 250

\bibitem[Krumholz 
\& Tan(2007)]{KT07} Krumholz, M.~R., \& Tan, J.~C.\ 2007, \apj, 654, 304 

\bibitem[Krumholz et al.(2012)]{KRU12} Krumholz, M.~R., 
Klein, R.~I., \& McKee, C.~F.\ 2012, arXiv:1203.2620 

\bibitem[Latif et al.(2011)]{LAT11a} Latif, M.~A., Zaroubi, 
S., \& Spaans, M.\ 2011, \mnras, 411, 1659 

\bibitem[Latif et 
al.(2011)]{LAT11b} Latif, M.~A., Schleicher, D.~R.~G., Spaans, M., \& Zaroubi, S.\ 2011, \aap, 532, A66 



\bibitem[Mac Low 
\& Klessen(2004)]{MK04} Mac Low, M.-M., \& Klessen, R.~S.\ 2004, Rev. Mod. Phys, 76, 125 

\bibitem[Machacek et al.(2001)]{MAC01} Machacek, M.~E., 
Bryan, G.~L., \& Abel, T.\ 2001, \apj, 548, 509

\bibitem[Mackey et al.(2003)]{MAC03} Mackey, J., Bromm, V., 
\& Hernquist, L.\ 2003, \apj, 586, 1 

\bibitem[Maio et al.(2010)]{MAI10} Maio, U., Ciardi, B., 
Dolag, K., Tornatore, L., \& Khochfar, S.\ 2010, \mnras, 407, 1003

\bibitem[McGreer 
\& Bryan(2008)]{MCG08} McGreer, I.~D., \& Bryan, G.~L.\ 2008, \apj, 685, 8

\bibitem[McKee 
\& Tan(2008)]{MT08} McKee, C.~F., \& Tan, J.~C.\ 2008, \apj, 681, 771 

\bibitem[McKee
\& Ostriker(2007)]{MO07} McKee, C.~F., \& Ostriker, E.~C.\ 2007, ARA\&A, 45, 565 

\bibitem[Mesinger et al.(2006)]{MES06} Mesinger, A., Bryan, 
G.~L., \& Haiman, Z.\ 2006, \apj, 648, 835 

\bibitem[Molina et al.(2012)]{MOL12} Molina, F.~Z., Glover, 
S.~C.~O., Federrath, C., \& Klessen, R.~S.\ 2012, arXiv:1203.2117

\bibitem[Nelson 
\& Langer(1997)]{NL97} Nelson, R.~P., \& Langer, W.~D.\ 1997, \apj, 482, 796 

\bibitem[Oh
\& Haiman(2002)]{OH02} Oh, S.~P., \& Haiman, Z.\ 2002, \apj, 569, 558

\bibitem[Oh 
\& Haiman(2003)]{OH03} Oh, S.~P., \& Haiman, Z.\ 2003, \mnras, 346, 456 

\bibitem[Omukai(2001)]{OMU01} Omukai, K.\ 2001, \apj, 546, 
635 

\bibitem[Omukai et al.(2005)]{OMU05} Omukai, K., Tsuribe, T.,
Schneider, R., \& Ferrara, A.\ 2005, \apj, 626, 627

\bibitem[Omukai et al.(2008)]{OMU08} Omukai, K., Schneider, 
R., \& Haiman, Z.\ 2008, \apj, 686, 801 

\bibitem[O'Shea et al.(2005)]{OSH05} O'Shea, B.~W., Abel, T., 
Whalen, D., \& Norman, M.~L.\ 2005, \apjl, 628, L

\bibitem[O'Shea 
\& Norman(2008)]{OSH08} O'Shea, B.~W., \& Norman, M.~L.\ 2008, \apj, 673, 14 

\bibitem[O'Shea et al.(2008)]{OSH08a} O'Shea, B.~W., McKee, 
C.~F., Heger, A., \& Abel, T.\ 2008, First Stars III, 990, 13

\bibitem[Osterbrock 
\& Ferland(2006)]{OF06} Osterbrock, D.~E., \& Ferland, G.~J.\ 2006, Astrophysics of gaseous nebulae and active galactic nuclei (2nd ed.;~Sausalito, CA: University Science Books)  

\bibitem[Ostriker et al.(2001)]{OS01} Ostriker, E.~C., 
Stone, J.~M., \& Gammie, C.~F.\ 2001, \apj, 546, 980

\bibitem[Padoan et al.(1997)]{PAD97} Padoan, P., Nordlund, 
A., \& Jones, B.~J.~T.\ 1997, \mnras, 288, 145 

\bibitem[Padoan 
\& Nordlund(2002)]{PN02} Padoan, P., \& Nordlund, {\AA}.\ 2002, \apj, 576, 870 

\bibitem[Padoan et al.(2007)]{PAD07} Padoan, P., Nordlund, 
{\AA}., Kritsuk, A.~G., Norman, M.~L., \& Li, P.~S.\ 2007, \apj, 661, 972 

\bibitem[Padoan 
\& Nordlund(2011)]{PN11} Padoan, P., \& Nordlund, {\AA}.\ 2011, \apj, 730, 40 

\bibitem[Palla et al.(1983)]{PAL83} Palla, F., Salpeter, 
E.~E., \& Stahler, S.~W.\ 1983, \apj, 271, 632

\bibitem[Pan et al.(2012)]{PAN12} Pan, T., Kasen, D., 
\& Loeb, A.\ 2012, \mnras, 2809 

\bibitem[Passot 
\& V{\'a}zquez-Semadeni(1998)]{PAS98} Passot, T., \& V{\'a}zquez-Semadeni, E.\ 1998, \pre, 58, 4501 

\bibitem[Pawlik et al.(2011)]{PAW11} Pawlik, A.~H., 
Milosavljevi{\'c}, M., \& Bromm, V.\ 2011, \apj, 731, 54 

\bibitem[Press 
\& Schechter(1974)]{PS74} Press, W.~H., \& Schechter, P.\ 1974, \apj, 187, 425

\bibitem[Price 
\& Monaghan(2007)]{PM07} Price, D.~J., \& Monaghan, J.~J.\ 2007, \mnras, 374, 1347 

\bibitem[Price et al.(2011)]{PRI11b} Price, D.~J., Federrath, 
C., \& Brunt, C.~M.\ 2011, \apjl, 727, L21

\bibitem[Prieto et al.(2012)]{PRI11} Prieto, J., Jimenez, R., 
\& Mart{\'{\i}}, J.\ 2012, \mnras, 419, 3092 

\bibitem[Prieto et al.(2011)]{PRI11a} Prieto, J., Padoan, P., 
Jimenez, R., \& Infante, L.\ 2011, \apjl, 731, L38 

\bibitem[Prunet et al.(2008)]{PRU08} Prunet, S., Pichon, C., 
Aubert, D., et al.\ 2008, \apjs, 178, 179

\bibitem[Regan 
\& Haehnelt(2009)]{RH09} Regan, J.~A., \& Haehnelt, M.~G.\ 2009, \mnras, 393, 858 

\bibitem[Regan 
\& Haehnelt(2009)]{RH09A} Regan, J.~A., \& Haehnelt, M.~G.\ 2009, \mnras, 396, 343

\bibitem[Ricker(2008)]{RIC08} Ricker, P.~M.\ 2008, \apjs, 
176, 293 

\bibitem[Ricotti et al.(2001)]{RIC01} Ricotti, M., Gnedin, 
N.~Y., \& Shull, J.~M.\ 2001, \apj, 560, 580 
 
\bibitem[Ritter et al.(2012)]{RIT12} Ritter, J.~S., 
Safranek-Shrader, C., Gnat, O., Milosavljevic, M., 
\& Bromm, V.\ 2012, arXiv:1203.2957 

\bibitem[Rydberg et al.(2010)]{RYD10} Rydberg, C.~E., 
Zackrisson, E., 
\& Scott, P.\ 2010, Cosmic Radiation Fields: Sources in the early Universe (CRF 2010), 2

\bibitem[Safranek-Shrader et al.(2010)]{SS10} 
Safranek-Shrader, C., Bromm, V., 
\& Milosavljevi{\'c}, M.\ 2010, \apj, 723, 1568 

\bibitem[Scannapieco et al.(2002)]{SCA02} Scannapieco, E., 
Ferrara, A., \& Madau, P.\ 2002, \apj, 574, 590

\bibitem[Scalo et al.(1998)]{SC98} Scalo, J., 
Vazquez-Semadeni, E., Chappell, D., \& Passot, T.\ 1998, \apj, 504, 83

\bibitem[Schaerer(2002)]{SCH02} Schaerer, D.\ 2002, \aap, 382, 28 

\bibitem[Schneider et 
al.(2012)]{SCH12} Schneider, N., Csengeri, T., Hennemann, M., et al.\ 2012, \aap, 540, L11 

\bibitem[Shang et al.(2010)]{SHA10} Shang, C., Bryan, G.~L., 
\& Haiman, Z.\ 2010, \mnras, 402, 1249

\bibitem[Shapiro \& Kang(1987)]{SK87}Shapiro, P.R., \& Kang, H. 1987,
ApJ, 318, 32

\bibitem[Sheth 
\& Tormen(2002)]{ST02} Sheth, R.~K., \& Tormen, G.\ 2002, \mnras, 329, 61 

\bibitem[Solomon 
\& Woolf(1973)]{SW73} Solomon, P.~M., \& Woolf, N.~J.\ 1973, \apjl, 180, L89

\bibitem[Stacy et al.(2010)]{STA10} Stacy, A., Greif, T.~H., 
\& Bromm, V.\ 2010, \mnras, 403, 45

\bibitem[Stacy et al.(2012)]{STA11} Stacy, A., Greif, T.~H., 
\& Bromm, V.\ 2012, \mnras, 422, 290


\bibitem[Stahler 
\& Palla(2005)]{SP05} Stahler, S.~W., \& Palla, F.\ 2005, The Formation of Stars (Weinheim, Germany: Wiley-VCH)

\bibitem[Stecher
\& Williams(1967)]{SW67} Stecher, T.~P., \& Williams, D.~A.\ 1967, \apjl, 149, L29

\bibitem[Stiavelli 
\& Trenti(2010)]{ST10} Stiavelli, M., \& Trenti, M.\ 2010, \apjl, 716, L190 

\bibitem[Tanaka et al.(2012)]{TAN12} Tanaka, M., Moriya, 
T.~J., Yoshida, N., \& Nomoto, K.\ 2012, \mnras, 2797 

\bibitem[Tegmark et al.(1997)]{TEG97} Tegmark, M., Silk, J.,
Rees, M.~J., Blanchard, A., Abel, T., \& Palla, F.\ 1997, \apj, 474, 1

\bibitem[Toomre(1964)]{TOM64} Toomre, A.\ 1964, \apj, 139, 
1217 

\bibitem[Tornatore et al.(2007)]{TOR07} Tornatore, L., 
Ferrara, A., \& Schneider, R.\ 2007, \mnras, 382, 

\bibitem[Trenti
\& Stiavelli(2009)]{TS09} Trenti, M., \& Stiavelli, M.\ 2009, \apj, 694, 879

\bibitem[Trenti et al.(2009)]{TRE09} Trenti, M., Stiavelli, 
M., \& Michael Shull, J.\ 2009, \apj, 700, 1672 

\bibitem[Truelove et al.(1997)]{TRU97} Truelove, J.~K., 
Klein, R.~I., McKee, C.~F., et al.\ 1997, \apjl, 489, L179

\bibitem[Turk et al.(2009)]{TUR09} Turk, M.~J., Abel, T.,
\& O'Shea, B.\ 2009, Science, 325, 601

\bibitem[Turk et al.(2012)]{TUR12} Turk, M.~J., Oishi, J.~S., 
Abel, T., \& Bryan, G.~L.\ 2012, \apj, 745, 154 

\bibitem[Vazquez-Semadeni(1994)]{VS94} Vazquez-Semadeni, E.\ 
1994, \apj, 423, 681 

\bibitem[Wang et al.(2010)]{WA10} Wang, P., Li, Z.-Y., Abel, 
T., \& Nakamura, F.\ 2010, \apj, 709, 27

\bibitem[Wise 
\& Abel(2005)]{WA05} Wise, J.~H., \& Abel, T.\ 2005, \apj, 629, 615 

\bibitem[Wise 
\& Abel(2007)]{WA07A} Wise, J.~H., \& Abel, T.\ 2007, \apj, 671, 1559 

\bibitem[Wise
\& Abel(2007)]{WA07} Wise, J.~H., \& Abel, T.\ 2007, \apj, 665, 899

\bibitem[Wise et al.(2008)]{WA08} Wise, J.~H., Turk, M.~J., 
\& Abel, T.\ 2008, \apj, 682, 745 

\bibitem[Wise et al.(2012)]{WIS12} Wise, J.~H., Turk, M.~J., 
Norman, M.~L., \& Abel, T.\ 2012, \apj, 745, 50

\bibitem[Wolcott-Green 
\& Haiman(2011)]{WG111} Wolcott-Green, J., \& Haiman, Z.\ 2011, \mnras, 412, 2603

\bibitem[Wolcott-Green et al.(2011)]{WG11} Wolcott-Green, 
J., Haiman, Z., \& Bryan, G.~L.\ 2011, \mnras, 418, 838 

\bibitem[Yoshida et al.(2006)]{Y06}Yoshida, N., Omukai, K.,
Hernquist, L., \& Abel, T. 2006, ApJ, 652, 6

\bibitem[Yoshida et al.(2007)]{YO07} Yoshida, N., Oh, S.~P., 
Kitayama, T., \& Hernquist, L.\ 2007, \apj, 663, 687 

\bibitem[Zuckerman 
\& Evans(1974)]{ZUC74} Zuckerman, B., \& Evans, N.~J., II 1974, \apjl, 192, L149 

\bibitem[Zackrisson et al.(2011)]{ZAK11} Zackrisson, E., 
Rydberg, C.-E., Schaerer, D., {\"O}stlin, G., 
\& Tuli, M.\ 2011, \apj, 740, 13

\bibitem[Zackrisson et al.(2012)]{ZAK12} Zackrisson, E., 
Zitrin, A., Trenti, M., et al.\ 2012, arXiv:1204.0517 

\bibitem[Zheng et al.(2012)]{ZH12} Zheng, W., Postman, M., 
Zitrin, A., et al.\ 2012, arXiv:1204.2305 

\end{thebibliography}
\end{document}